\documentclass[11pt,letterpaper]{article}
\usepackage{jheppub}
\usepackage{graphicx}
\input{epsf}
\usepackage{epsfig}
\usepackage{epstopdf}
\usepackage{arcs}
\usepackage{subfigure}
\usepackage{tensor}
\usepackage{braket}
\usepackage{amsmath}
\newcommand{\be}{\begin{equation}}
\newcommand{\ee}{\end{equation}}

\usepackage{tikz}
\usetikzlibrary{calc}   
\usetikzlibrary{topaths}
\usepackage{pgfplots}
\newcommand{\labell}[1]{\label{#1}}

\newcommand{\bea}{\begin{eqnarray}}
\newcommand{\eea}{\end{eqnarray}}
\newcommand{\ba}{\begin{eqnarray}}
\newcommand{\ea}{\end{eqnarray}}

\newcommand{\beq}{\begin{equation}}
\newcommand{\eeq}{\end{equation}}
\newcommand{\beqa}{\begin{eqnarray}}
\newcommand{\eeqa}{\end{eqnarray}}
\newcommand{\beqar}{\begin{eqnarray*}}
\newcommand{\eeqar}{\end{eqnarray*}}
\newcommand{\e}{\epsilon}

\newcommand{\X}{\mathcal{X}}
\newcommand{\W}{\mathcal{W}}
\newcommand{\R}{\mathcal{R}}
\newcommand{\Z}{\mathcal{Z}}
\newcommand{\A}{\mathcal{A}}

\newcommand{\C}{\mathcal{C}}
\newcommand{\D}{\mathcal{D}}
\newcommand{\E}{\mathcal{E}}
\newcommand{\G}{\mathcal{G}}

\newcommand{\dslash}{\delta^{\!\!\!\!-}\!}

\renewcommand{\c}{$c$}

\renewcommand{\href}[2]{#2}

\title{Lovelock action with nonsmooth boundaries}
\author[a,b]{Pablo A. Cano}

\affiliation[a]{Instituto de F\'isica Te\'orica UAM/CSIC,\\C/ Nicol\'as Cabrera, 13-15, C.U. Cantoblanco, 28049 Madrid, Spain}
\affiliation[b]{Perimeter Institute for Theoretical Physics, Waterloo, ON N2L 2Y5, Canada}

\vspace{0.2cm}

\emailAdd{pablo.cano@uam.es}

\abstract{We examine the variational problem in Lovelock gravity when the boundary contains timelike and spacelike segments nonsmoothly glued. We show that two kinds of contributions have to be added to the action. The first one is associated to the presence of a boundary in every segment and it depends on intrinsic and extrinsic curvatures. We can think of this contribution as adding a total derivative to the usual surface term of Lovelock gravity. The second one appears in every joint between two segments and it involves the integral along the joint of the Jacobson-Myers entropy density weighted by the Lorentz boost parameter which relates the orthonormal frames in each segment.  We argue that this term can be straightforwardly extended to the case of joints involving null boundaries. As an application, we compute the contribution of these terms to the complexity of global AdS in Lovelock gravity by using the ``complexity = action'' proposal and we identify possible universal terms for arbitrary values of the Lovelock couplings. We find that they depend on the charge $a^*$ controlling the holographic entanglement entropy and on a new constant that we characterize. 
}

\preprint{IFT-UAM/CSIC-18-020
}

\begin{document}
\maketitle

\section{Introduction}
It has been known for a very long time that the gravitational action needs to be supplemented with boundary terms in order for it to define a well-posed variational problem \cite{York:1972sj,Gibbons:1976ue}. Well-posedness means that the solution of the equations of motion with some fixed boundary conditions must be the only extremum of the action when we perform variations that keep fixed the boundary data \cite{Dyer:2008hb}.  Although the surface terms do not modify the equations of motion, they play a crucial role in the Hamiltonian formalism \cite{Hawking:1995fd} or if we want to define a partition function for gravity \cite{Gibbons:1976ue}, something which is particularly relevant, for example, in the context of holography \cite{Maldacena, Witten, Gubser}. In the case of Einstein gravity, the appropriate surface contribution for spacelike and timelike boundaries is the well-known York-Gibbons-Hawking (YGH) term \cite{York:1972sj,Gibbons:1976ue}, which involves the integral over the boundary of the trace of its extrinsic curvature. However, situations more general than spacelike or timelike boundaries may appear. For example, the YGH term ensures the well-posedness of the variational principle when the boundary is smooth, but in certain cases the boundary may contain corners --- joints between different segments of the boundary where there is a discontinuity in the normal vector. In such cases, additional terms have to be added to the action in order to account for the nonsmoothness of the boundary\footnote{In the mathematical literature, these terms have been studied in the context of the Gauss-Bonnet theorem, see {\it e.g.} \cite{Jee1984,10.2307/44237512,birman1984} } \cite{Hartle:1981cf,Hayward:1993my}. These joints appear naturally in some situations, \textit{e.g.} when computing the Euclidean action of certain configurations \cite{Brill:1991rw, Hawking:1994ii,Gibbons:1994ff,FARHI1990417} or when defining a quasi-local energy of the gravitational field in a spatially bounded region \cite{BROWN2002175,PhysRevD.47.1407}. A more recent motivation comes from the ``complexity=action'' proposal \cite{Brown:2015bva,Brown:2015lvg} in the context of holography, which involves the computation of the gravitational action in the so-called Wheeler-DeWitt (WDW) patch of asymptotically Anti-de Sitter spaces \cite{Carmi:2016wjl}. Besides containing joints, the WDW patch is delimited by null boundaries, where the standard YGH surface term is not applicable. Fortunately, the null boundary terms for Einstein gravity were recently described \cite{Neiman:2012fx, Parattu:2015gga}, but it was found that these terms present ambiguities associated to the freedom to choose the parametrization of the null generators. The definitive step came in \cite{Lehner:2016vdi}, where the complete gravitational action with all kind of boundaries and joints was studied and also a prescription to cure the ambiguities of the null boundaries --- by demanding additivity of the action --- was introduced.\footnote{See also \cite{Jubb:2016qzt} for a revisited computation of the gravitational action.}

Much less is known about surface terms in the case of higher-derivative gravity. Several generalizations of the YGH term exist for some theories, \textit{e.g.} \cite{Smolic:2013gz,Madsen:1989rz,Dyer:2008hb,Guarnizo:2010xr,Love,Deruelle:2009zk,Teimouri:2016ulk,Bueno:2018xqc}, but the variational problem is not fully understood in general because these theories usually contain additional degrees of freedom, \textit{e.g.} \cite{Prue,Tekin1,Aspects}. As a consequence, it is unclear which variables one should keep fixed on the boundary. An exception to this is Lovelock gravity \cite{Lovelock1,Lovelock2}, which is the most general higher-curvature theory of gravity whose equations of motion are of second-order. This crucial property ensures that it is possible to obtain a well-posed variational principle for Lovelock gravity upon the addition of some appropriate boundary terms. In the case of spacelike and timelike boundaries, the surface terms were independently constructed by Myers \cite{Myers:1987yn} and Teitelboim and Zanelli \cite{Teitelboim:1987zz} --- we will review them in section \ref{21}. However, there is still work to be done in order to understand Lovelock variational principle in the most general region: the surface terms for null boundaries are not yet known, and the contribution from joints is also unknown for any kind of boundary.

As a step forward into comprehending Lovelock's action in the most general case, in this work we compute the joint terms when the boundary contains spacelike and timelike segments. However, we will see that an important part of the result is clearly generalizable to the case of joints involving null segments. 

The paper is organized as follows. Next we summarize how to compute the action in Lovelock gravity in the presence of joints, while the detailed derivation of this result is addressed in section \ref{2}. In subsection \ref{21} we review the surface terms in Lovelock gravity. In subsections \ref{22} and \ref{23} we compute the contribution of timelike joints and of spacelike joints of a special type by using the smoothing method of \cite{Hayward:1993my}. In subsection \ref{24} we show how to generalize these contributions to all kind of joints, and even to joints involving null boundaries. In section \ref{3} we explore the consequences of this result for holographic complexity of global AdS in Lovelock gravity. We compute the contribution to the complexity from the joints and from the bulk of the WDW patch and we identify universal terms in the cutoff expansion. Although the null surface terms are not yet known, we argue that probably they will not change this result. We discuss the results obtained in section \ref{4}.

\subsection{The complete Lovelock action}\label{CLove}
Let $\mathcal{M}$ be a pseudo-Riemannian manifold whose boundary is composed of nonsmoothly glued segments $\partial \mathcal{M}=\cup_{k}\mathcal{B}_k$ which we allow to be spacelike or timelike, but not null. The intersection of two of these segments is a codimension 2 surface that we denote by $\mathcal{C}_l=\mathcal{B}_{k_1}\cap \mathcal{B}_{k_2}$ and where there is a discontinuity in the normal vector. Alternatively, we can think of $\mathcal{C}_l$ as the common boundary of these segments $\partial \mathcal{B}_{k_1}=\partial \mathcal{B}_{k_2}$.  In $D=d+1$ dimensions there are $\lfloor D/2\rfloor$ independent terms that can be added to the Lovelock action, which in general will be a linear combination of the form $I=\sum_{n=1}^{\lfloor D/2\rfloor}\lambda_n I^{(n)}$. Then, the variational problem is well-posed if the $n$-th action $I^{(n)}$ is given by
\begin{equation}\label{LoveFull}
I^{(n)}=\int_{\mathcal{M}}d^{d+1}x\sqrt{|g|}\mathcal{X}_{2n}+\sum_{k}\left[\int_{\mathcal{B}_k}d\Sigma \mathcal{Q}_{n}+\int_{\partial\mathcal{B}_k}d\sigma \mathcal{F}_{n}\right]+\sum_l\int_{\mathcal{C}_l}d\sigma 2n\psi \hat{\mathcal{X}}_{2(n-1)}\, .
\end{equation}

Let us explain every term in this expression
\begin{itemize}
\item{$\mathcal{X}_{2n}$ is the dimensionally continued $2n$-dimensional Euler density, given by\footnote{The alternate Kronecker symbol is defined as: $\delta^{\mu_1\mu_2\dots \mu_r}_{\nu_1\nu_2\dots\nu_r}= r!\delta^{[\mu_1}_{\nu_1}\delta^{\mu_2}_{\nu_2}\dots \delta^{\mu_r]}_{\nu_r}$.
}
\begin{equation}
\mathcal{X}_{2n}= \frac{1}{2^{n}}\delta^{\mu_1\dots \mu_{2n}}_{\nu_1\dots \nu_{2n}}R^{\nu_1\nu_2}_{\mu_1\mu_2}\dots R^{\nu_{2n-1}\nu_{2n}}_{\mu_{2n-1}\mu_{2n}}\, .
\end{equation}}
\item{$\mathcal{Q}_{n}$ is the generalized York-Gibbons-Hawking boundary term, which is given by
\begin{equation}
\mathcal{Q}_{n}=2n\int_0^1dt\, \delta^{i_1\dots i_{2n-1}}_{j_1\dots j_{2n-1}}K^{j_1}_{i_1}\left[\frac{1}{2}\mathcal{R}^{j_2 j_3}_{i_2i_3}-\epsilon t^2K^{j_2}_{i_2}K^{j_3}_{i_3}\right]\cdots \left[\frac{1}{2}\mathcal{R}^{j_{2n-2}j_{2n-1}}_{i_{2n-2}i_{2n-1}}-\epsilon t^2K^{j_{2n-2}}_{i_{2n-2}}K^{j_{2n-1}}_{i_{2n-1}}\right]\, ,
\label{Qterm1}
\end{equation}
where $\mathcal{R}^{j_2 j_3}_{i_2i_3}$ is the intrinsic curvature of the corresponding boundary segment, $K^{i}_{j}$ is the extrinsic curvature and $\epsilon=n^2=\pm 1$ is the sign of the normal to the boundary. Also, $d\Sigma=d^dx\sqrt{|h|}$ is the volume element on $\mathcal{B}_k$ and the orientation is such that, as a 1-form, $n=n_{\mu}dx^{\mu}$ points outside of $\mathcal{M}$.
}

\item{There is also a contribution associated to the boundary $\partial\mathcal{B}_k$ of every segment. Let us introduce $s^i$ as the tangent vector in $\mathcal{B}_k$ which is normal to $\partial\mathcal{B}_k$ and let us introduce as well a basis of tangent vectors to $\partial\mathcal{B}_k$ as $e_A^i$, $A, B= 2,...,d$. Then let $Q_{AB}=e_{A}^ie_{B}^jD_i s_j$ be the extrinsic curvature of $\partial\mathcal{B}_k$ from the point of view of $\mathcal{B}_k$, where $D_i$ is the covariant derivative in $\mathcal{B}_k$. Also, $K_{AB}=e_{A}^i e_{B}^jK_{ij}$ is the projection of the extrinsic curvature of $\mathcal{B}_k$ onto its boundary and in the same way $R^{A_1A_2}_{B_1B_2}$ is the projection of the spacetime curvature. This can also be expressed in terms of the intrinsic and extrinsic curvatures by using 
\begin{equation}
R^{B_1B_2}_{A_1A_2}=\mathcal{R}^{B_1B_2}_{A_1A_2}-\frac{2}{n^2} K^{B_1}_{[A_1}K^{B_2}_{A_2]}-\frac{2}{s^2} Q^{B_1}_{[A_1}Q^{B_2}_{A_2]}\, ,
\end{equation}
where $n^2$ and $s^2$ are the norms of $n$ and $s$ respectively.
Then, in (\ref{LoveFull}) $d\sigma$ is the volume element in $\partial\mathcal{B}_k$ and the term $\mathcal{F}_n$ is
\begin{equation}
\mathcal{F}_n= \sum_{l=2}^{n}\frac{n!(l-1)! \epsilon^{l-1}R^{n-l}}{(n-l)!2^{n-l}}\sum_{0=j\neq l-1}^{2l-2}\frac{(K+Q)^j(K-Q)^{2l-2-j}}{j!(2l-2-j)!(l-1-j)}\, ,
\end{equation}
for spacelike joints and
\begin{equation}
\begin{aligned}
\mathcal{F}_n&=\sum_{l=2}^{n}\frac{n!(l-1)! R^{n-l}}{2^{n-l-1}(n-l)!}{\rm Im}\left[\sum_{j=0}^{l-2}\frac{(K-i Q)^j(K+i Q)^{2l-2-j}}{j!(2l-2-j)!(l-1-j)}\right]\, ,
\end{aligned}
\end{equation}
for timelike ones. In these expressions we are using the short-hand notation

\begin{equation}
K^{2l-2}R^{n-l}\equiv \delta^{A_2 \dots A_{2n-1}}_{B_2\dots B_{2n-1}}K^{B_2}_{A_2}\cdots K^{B_{2l-2}}_{A_{2l-2}}K^{B_{2l-1}}_{A_{2l-1}}R^{B_{2l} B_{2l+1}}_{A_{2l}A_{2l+1}}\cdots R^{B_{2n-2}B_{2n-1}}_{A_{2n-2}A_{2n-1}}\, .
\end{equation}
Note that although the contribution $\int_{\partial\mathcal{B}_k}d\sigma \mathcal{F}_{n}$ involves an integral over a joint, it actually does not depend on which other segment $\mathcal{B}_k$ is glued to, and therefore, it should be considered a part of the boundary term. Indeed, we may reinterpret this term as adding the total derivative $D_i\left(\frac{s^i}{s^2}\mathcal{F}_n\right)$ to $\mathcal{Q}_n$, for which we would need to extend the definition of $s^i$ to the interior of $\mathcal{B}_k$.\footnote{Similar terms have been obtained in the context of Lovelock theory with localized defects \cite{Gravanis:2003aq,Gravanis:2004kx,Charmousis:2005ey}. It would be interesting to further explore the relation between those terms and the ones introduced here.}}

\item{The contribution of the joint contains the $(n-1)$th Euler density $\hat{\mathcal{X}}_{2(n-1)}$ constructed with the curvature of the induced metric $\sigma_{AB}$ on $\mathcal{C}_l$\footnote{We use the convention $\hat{\mathcal{X}}_{0}=1$.}. The parameter $\psi$ measures the change in the normal at the joint, and the rules to assign it are the same as in Einstein gravity. A detailed analysis was carried out in \cite{Lehner:2016vdi}. In particular, for timelike joints $\psi=\Theta\equiv\arccos(n_1\cdot n_2)$ is the angle in which the normal changes, while for spacelike joints $\psi=\pm\eta$ is the rapidity parameter associated to the Lorentz boost which connects the orthonormal frames in $\mathcal{B}_{k_1}$ and $\mathcal{B}_{k_2}$. This term is also present in joints involving null boundaries and in such case $\psi$ also takes the same value as in Einstein gravity ($\psi$ is equal to the parameter $a$ introduced in \cite{Lehner:2016vdi})  }
\end{itemize}

For example, for Einstein-Gauss-Bonnet gravity the action (including only spacelike joints) reads explicitly
\begin{equation}
\begin{aligned}
I=&\int_{\mathcal{M}}d^{d+1}x\sqrt{|g|}\left[R+\lambda \mathcal{X}_4\right]\\
&+ \sum_{k}\Bigg\{2\int_{\mathcal{B}_k}d\Sigma\left[K+\lambda\delta^{i_1i_2i_3}_{j_1j_2j_3}K^{j_1}_{i_1}\left(\mathcal{R}^{j_2 j_3}_{i_2i_3}-\epsilon \frac{2}{3}K^{j_2}_{i_2}K^{j_3}_{i_3}\right)\right]-8\lambda\int_{\partial\mathcal{B}_k}d\sigma  \epsilon K^{[A}_{A}Q^{B]}_{B}\Bigg\}\\
&+\sum_l\int_{\mathcal{C}_l}d\sigma 2\eta \left(1+2\lambda\hat{\mathcal{R}}\right)\, .
\end{aligned}
\end{equation}

Note that the contribution from the joint contains the Jacobson-Myers entropy density \cite{Jacobson:1993xs}:
\begin{equation}\label{Ijnt1}
I_{\rm joint}=\int_{\mathcal{C}}d\sigma \frac{\psi}{2\pi} \rho_{\rm JM}\, ,
\end{equation}
where 
\begin{equation}
\rho_{\rm JM}=\sum_{n=1}4\pi n\lambda_n\hat{\mathcal{X}}_{2(n-1)}\, .
\end{equation}

\subsubsection*{Conventions}
The metric has mostly $+$ signature: $\operatorname{sign}{g}=(-,+,+,...,+)$. The space-time dimension is $D=d+1$.  We use Greek letters to denote spacetime indices $\mu, \nu=0, \ldots, d$, Latin letters $i, j$  to denote boundary indices and capital letters $A, B$ to denote indices on the joints $\mathcal{C}$.
The covariant derivative is defined by
\begin{equation}
\nabla_{\mu}u_{\nu}=\partial_{\mu}u_{\nu}-u_{\lambda}\Gamma^{\ \ \ \lambda}_{\mu\nu}\, .
\end{equation}
The curvature is defined by
\begin{equation}
u_{\mu}R^{\mu}_{\ \ \nu\rho\sigma}=-\left[\nabla_{\rho},\nabla_{\sigma}\right] u_{\nu}\, ,
\end{equation}
and similarly for the different intrinsic curvatures. In terms of the Christoffel symbols it reads
\begin{equation}
R^{\mu}_{\ \ \nu\rho\sigma}=\partial_{\rho}\Gamma_{\sigma\nu}^{\ \ \ \mu}-\partial_{\sigma}\Gamma_{\rho\nu}^{\ \ \ \mu}+\Gamma_{\rho\lambda}^{\ \ \ \mu}\Gamma_{\sigma\nu}^{\ \ \ \lambda}-\Gamma_{\sigma\lambda}^{\ \ \ \mu}\Gamma_{\rho\nu}^{\ \ \ \lambda}
\end{equation}

\section{Contribution of joints in the Lovelock action}\label{2}
In this section we derive the gravitational action (\ref{LoveFull}). In \ref{21} we review the surface terms for timelike and spacelike boundaries in Lovelock gravity and in \ref{22} and \ref{23} we compute the contribution from the joints by taking an appropriate limit in the surface term. This method is only applicable to some kinds of spacelike joints --- those that we will call of type I ---, so in \ref{24} we generalize the result to all kinds of joints. The method that we use to obtain the general result also gives us relevant information when the boundaries are null, so that we are able to derive the joint term (\ref{Ijnt1}) in that case as well.

\subsection{Variational problem in Lovelock gravity}\label{21}
Lovelock gravity in $D=d+1$ dimensions is given by the bulk action
\begin{equation}\labell{flov}
I_{\rm bulk}=\int_{\mathcal{M}}d^{d+1}x\sqrt{|g|}\sum_{n=0}^{\lfloor D/2 \rfloor}\lambda_n\mathcal{X}_{2n}=\sum_{n=0}^{\lfloor D/2 \rfloor} \lambda_n I_{\rm bulk}^{(n)}\, 
\end{equation}
where $\lambda_n$ are arbitrary constants and the dimensionally extended Euler densities (ED) $\mathcal{X}_{2n}$ are defined as 
\begin{equation}
\mathcal{X}_{2n}= \frac{1}{2^{n}}\delta^{\mu_1\dots \mu_{2n}}_{\nu_1\dots \nu_{2n}}R^{\nu_1\nu_2}_{\mu_1\mu_2}\dots R^{\nu_{2n-1}\nu_{2n}}_{\mu_{2n-1}\mu_{2n}}\, .
\end{equation}
The first cases are $\mathcal{X}_2=R$,  $\mathcal{X}_{4}=R^2-4R_{\mu\nu}R^{\mu\nu}+R_{\mu\nu\rho\sigma}R^{\mu\nu\rho\sigma}$, this is, the Ricci scalar and Gauss-Bonnet (GB) terms respectively. Note that $\mathcal{X}_{2n}$ vanishes identically for $n> \lfloor D/2 \rfloor$, and it is topological for $D=2n$.

Let us compute the variation of the $n$-th Lovelock action $I_{\rm bulk}^{(n)}$ with respect to the metric. Assuming that the space-time manifold $\mathcal{M}$ has spacelike or timelike boundaries, we find\footnote{As in \cite{Lehner:2016vdi}, we use the symbol $\dslash f $ to denote an infinitesimal quantity that is not actually the total variation of some other quantity $f$.}
\begin{equation}
\delta I_{\rm bulk}^{(n)}=\int_{\mathcal{M}} d^{d+1}x\sqrt{|g|}\,\E^{(n)}_{\mu\nu}\delta g^{\mu\nu}+\int_{\partial \mathcal{M}} d\Sigma_{\mu}\dslash v^{\mu}_{n}\, ,\quad \text{where}\, \, \, \,\dslash  v^{\mu}_{n}=2 P^{(n)\,\,\,\,\beta\mu\nu} _{\,\,\,\,\,\,\,\,\, \alpha}\delta \Gamma^{\ \ \ \alpha}_{\nu\beta}\, ,
\label{varf}
\end{equation}
where we have used Stokes' theorem in the second term. Here, the volume element of the boundary is $d\Sigma_{\mu}=d^{d}x\sqrt{|h|} n_{\mu}$, where $n_{\mu}$ is the outward-directed normal 1-form to the boundary $\partial \mathcal{M}$ with $n_{\mu}n^{\mu}=\epsilon=\pm 1$. Note that this implies that for spacelike boundaries the normal vector $n^{\mu}$ is inward-directed: it points to the future when the boundary is in the past of $\mathcal{M}$ and vice versa \cite{Carroll:2004st}. Also, the induced metric on $\partial\mathcal{M}$ is given by $h_{\mu\nu}=g_{\mu\nu}-\epsilon\, n_{\mu}n_{\nu}$, and we have introduced the tensors
\begin{equation}
\mathcal{E}_{\mu\nu}^{(n)}=\frac{-1}{2^{n+1}}g_{\alpha \mu}\delta^{\alpha \mu_1\dots \mu_{2n}}_{\nu \nu_1\dots \nu_{2n}} R^{\nu_1\nu_2}_{\mu_1\mu_2}\cdots R^{\nu_{2n-1}\nu_{2n}}_{\mu_{2n-1}\mu_{2n}}\, , \quad P^{(n)\mu\nu}_{\alpha\beta}=\frac{n}{2^n}\delta^{\mu\nu \sigma_1\dots\sigma_{2n-2}}_{\alpha \beta \lambda_1\dots \lambda_{2n-2}}R_{\sigma_1\sigma_2}^{\lambda_1\lambda_2}\cdots R_{\sigma_{2n-3}\sigma_{2n-2}}^{\lambda_{2n-3}\lambda_{2n-2}} \, .
\label{love}
\end{equation}
Note that the equations of motion for Lovelock gravity are
\begin{equation}
\sum_{n=0}^{\lfloor D/2 \rfloor} \lambda_n\mathcal{E}_{\mu\nu}^{(n)}=0\, ,
\end{equation}
which are of second order in derivatives of the metric. In order to work with the surface terms it is useful to introduce a basis of tangent vectors ${e_i^{\mu}}$ in the boundary $\partial\mathcal{M}$. These satisfy $e_i^{\mu}n_{\mu}=0$ and we can write the induced metric on the boundary as
\begin{equation}
h_{ij}=e_i^{\mu}e_j^{\nu}g_{\mu\nu}=e_i^{\mu}e_j^{\nu}h_{\mu\nu}\, .
\end{equation}
Now, in order to have a well-posed variational problem we must demand that the action is stationary around solutions of the equations of motion for variations satisfying $\delta h_{ij}=0$.  Note that this does not imply $\delta g_{\mu\nu}\Big|_{\partial \mathcal{M}}=0$, nor $\nabla\delta g_{\mu\nu}\Big|_{\partial \mathcal{M}}=0$, so the variational problem (\ref{varf}) is not well-posed. It is known that for spacelike or timelike boundaries the Lovelock action becomes well-posed if one adds to it the following boundary contribution \cite{Teitelboim:1987zz,Myers:1987yn}:

\begin{equation}\label{bdry}
I_{\rm bdry}^{(n)}=\int_{\partial \mathcal{M}}d^{d}x\sqrt{|h|}\mathcal{Q}_{n}\, ,
\end{equation}
where
\begin{equation}
\mathcal{Q}_{n}=2n\int_0^1dt\, \delta^{i_1\dots i_{2n-1}}_{j_1\dots j_{2n-1}}K^{j_1}_{i_1}\left[\frac{1}{2}\mathcal{R}^{j_2 j_3}_{i_2i_3}-\epsilon t^2K^{j_2}_{i_2}K^{j_3}_{i_3}\right]\cdots \left[\frac{1}{2}\mathcal{R}^{j_{2n-2}j_{2n-1}}_{i_{2n-2}i_{2n-1}}-\epsilon t^2K^{j_{2n-2}}_{i_{2n-2}}K^{j_{2n-1}}_{i_{2n-1}}\right]\, ,
\label{Qterm}
\end{equation}
and where $\mathcal{R}^{j_1j_2}_{i_1i_2}$ is the curvature of the induced metric $h_{ij}$ and $K_{ij}$ is the extrinsic curvature of the boundary, defined as
\begin{equation}
K_{ij}=e^{\mu}_{i}e^{\nu}_{j}\nabla_{\mu}n_{\nu}=\frac{1}{2}\mathcal{L}_{n} h_{ij}\, .
\end{equation}
Different derivations of this result can be found in the literature \cite{Chakraborty:2017zep}, but for the sake of completeness, in appendix A we show with a direct computation that when this boundary term is added to the action, the total variation reads
\begin{equation}\label{var}
\int_{\partial \mathcal{M}} d\Sigma_{\mu}\dslash  v^{\mu}_{n}+\delta I_{\rm bdry}^{(n)}=\int_{\partial\mathcal{M}} d^{d}x\sqrt{|h|}\left( T^{ij}\delta h_{ij}+D_i \dslash H^i\right)\, ,
\end{equation}
for certain $T^{ij}$ that we do not worry about, and the expression for $\dslash H^i$ can be found in the appendix. The boundary as a whole is a closed hypersurface, so when it is smooth the total derivative terms vanish and the variational problem is well-posed. However, if the boundary is composed of several pieces nonsmoothly glued, these terms play a role, as they contribute differently in every segment. By using Stokes' theorem again we may rewrite $D_i \dslash H^i$ as an integral over the boundary of every segment --- a joint ---, and the task would then be to express this contribution as the variation of a quantity defined on the joint. Then, we must subtract this quantity in the action in order to obtain a well-posed variational problem. However, this process is considerably non-straightforward, and in order to obtain these corner or joint contributions we may use a different method. A possible approach, first used by Hayward \cite{Hayward:1993my}, consists in considering a smoothed version of the boundary, in which no corner terms are necessary, and at the end take the limit in which the boundary becomes sharp. In the case of Euclidean signature this method works for any joint, but in Lorentzian signature it has the disadvantage that it can only be applied to certain kinds of joints.  We distinguish between type I joints, which can be replaced by an smooth boundary in which the normal vector interpolates continuously between one side and the other of the joint, and type II joints for which this is not possible, because the normal would become null at certain point. Hence, the smoothing procedure only works for those of type I and we should use the variational method for type II joints. In the two next sections we are going to use the smoothing method in order to determine the contribution from type I joints, but afterwards, in section \ref{24}, we will see that the result can be straightforwardly generalized for type II joints as well.


\subsection{Timelike joints}\label{22}
\begin{figure}[t]
  \centering
   \includegraphics[height=5.5cm]{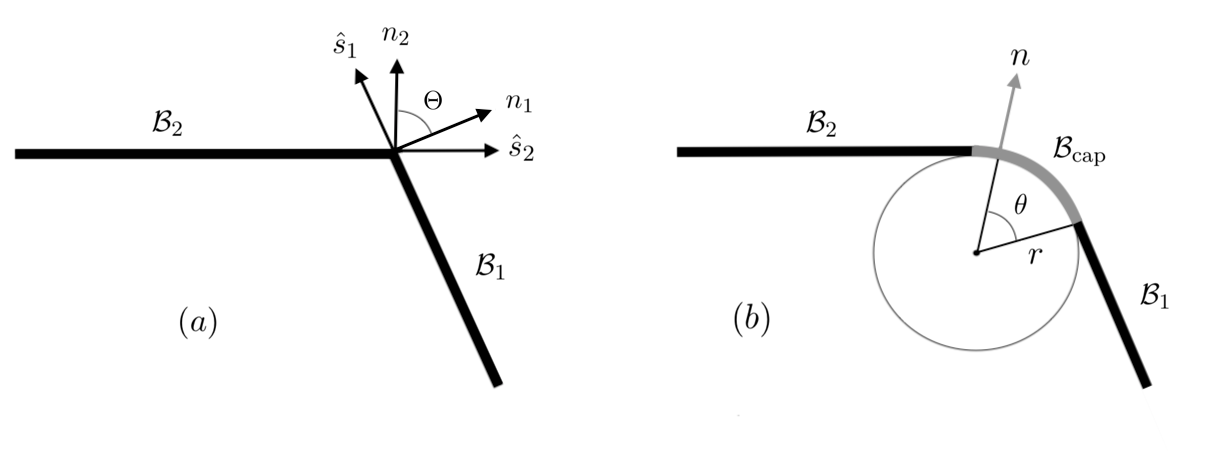}
  \caption{\small Timelike joints. (a) We show the normal 1-forms $n=n_{\mu}dx^{\mu}$ and the tangent vectors $\hat s=s^{\mu}\partial_{\mu}$ at the joint. (b) We replace the joint by a smooth cap of certain radius $r$. The joint is recovered in the limit $r\rightarrow 0$.}
  \label{Tmlj}
\end{figure}

Let us consider the case in which the joint has two spacelike normals. This is always the case for Euclidean signature, while for Lorentzian signature we say that the joint is a timelike codimension 2 surface, since it contains a timelike tangent vector. Let $\mathcal{B}_1$ and $\mathcal{B}_2$ be the segments of boundary that intersect at the joint $\mathcal{C}=\mathcal{B}_1\cap \mathcal{B}_2$. Let $n_1$ and $n_2$ be the normal 1-forms in each segment and let us define $\Theta= \arccos(n_1\cdot n_2)$ as the angle in which the normal changes at the joint. It will be useful to introduce as well the vectors $\hat s_1$ and $\hat s_2$\footnote{We use hats to distinguish vectors $\hat s=s^{\mu}\partial_{\mu}$ from 1-forms $s=s_{\mu}dx^{\mu}$. Note that in this case we choose the vector $\hat s$ --- not the 1-form --- to be pointing outwards. For timelike joints this makes no difference but it will be relevant for spacelike joints.} which are tangent to $\mathcal{B}_1$ and $\mathcal{B}_2$ respectively and which are normal to $\mathcal{C}$ pointing outwards their respective segment --- see Figure \ref{Tmlj} (a). Hence, the orthonormal systems at the joint will be related according to
\begin{equation}\label{rot}
\begin{aligned}
n_2&=n_1 \cos\Theta+s_1 \sin \Theta \, ,\\
s_2&=n_1 \sin\Theta -s_1 \cos \Theta\, 
\end{aligned}
\end{equation}

Then, following \cite{Hayward:1993my} we are going to replace the joint by a cap of certain size $r$, apply the boundary term (\ref{bdry}) to this smoothed boundary and then take the limit in which the cap becomes a sharp corner, this is, $r\rightarrow 0$. The smoothed boundary can be split in two parts: $\partial\mathcal{M}=\left(\mathcal{B}_1\cup\mathcal{B}_2-\mathcal{B}_{\rm cap}\right)\cup \mathcal{B}_{\rm cap}$, so that 

\begin{equation}
I_{\rm bdry}^{(n)}=\int_{\mathcal{B}_1\cup\mathcal{B}_2-\mathcal{B}_{\rm cap}}d^{d}x\sqrt{|h|}\mathcal{Q}_{n}+\int_{\mathcal{B}_{\rm cap}}d^{d}x\sqrt{|h|}\mathcal{Q}_{n}\, .
\end{equation}
The first integral involves an smooth surface when we take the size of the cap to zero. Let us then evaluate the second integral. The easiest way to proceed is to consider a locally Gaussian coordinate frame, in which we can always choose the metric to have the form\footnote{Actually, if we only consider a thin slice of $\mathcal{C}$ we can always assume that near the joint the metric is locally Minkowskian.}
\begin{equation}
ds^2=n^2+h_{ij}dx^{i}dx^{j}\, ,
\end{equation} 
with
\begin{equation}
n=N dr\, ,\quad h_{ij}dx^{i}dx^{j}=M^2 d\tilde\theta^2+\sigma_{AB}dx^{A}dx^{B}\, .
\end{equation} 
Here $r$ represents a polar ``radial" coordinate from certain axis and $\tilde\theta$ represents rotation around this axis ---see Figure \ref{Tmlj} (b). Note that $n$ is the normal to $\mathcal{B}_{\rm cap}$, while $h_{ij}$ is the induced metric. A set of constraints is obtained by demanding regularity of the metric at the axis $r=0$:
\begin{equation}
M\Big|_{r\rightarrow 0} =M(r)\, , \quad M\Big|_{r= 0} =0\, ,\quad \frac{\partial N}{\partial\tilde\theta}\Big|_{r=0}=\frac{\partial \sigma_{AB}}{\partial\tilde\theta}\Big|_{r=0}=0\, .
\end{equation}
In addition, in order to avoid a conical singularity we must have

\begin{equation}
\lim_{r\rightarrow 0}\frac{\partial_r M}{N}=1\, .
\end{equation}
Now, in this coordinate frame the extrinsic curvature of the cap takes the particularly simple form
\begin{equation}
K_{ij}=\frac{1}{2N}\partial_r h_{ij}\, .
\end{equation}
Raising one index with $h^{jk}$, the non-vanishing components are, 
\begin{equation}
K^{\tilde \theta}_{\tilde \theta}=\frac{\partial_r M}{M N}\, ,\quad K^{B}_{A}=\frac{1}{2N}\partial_r \sigma_{AC}\sigma^{CB}\, .
\end{equation}
Note that $K^{B}_A$ is actually the extrinsic curvature of the joint $\mathcal{C}$ associated to the normal $n$: $K_{AB}=\frac{1}{2}\mathcal{L}_{n}\sigma_{AB}$.
On the other hand, the component  $K^{\tilde \theta}_{\tilde \theta}$ diverges as $1/M$ in the limit $r\rightarrow 0$. However, the volume element reads $\sqrt{|h|}=M\sqrt{|\sigma|}$, so that it goes to zero in that limit. Therefore, only the terms linear in $K^{\tilde \theta}_{\tilde \theta}$ will give a non-vanishing contribution. Terms with more than one $K^{\tilde \theta}_{\tilde \theta}$ would be divergent, but there are not such terms due to the antisymmetric character of the boundary contribution (\ref{Qterm}).
Before taking the limit $r\rightarrow 0$ in (\ref{bdry}), let us rewrite the intrinsic curvature in terms of the spacetime curvature and of the extrinsic curvature, so that (\ref{Qterm}) takes the form
\begin{equation}
\mathcal{Q}_{n}=2n\int_0^1dt\, \delta^{i_1\dots i_{2n-1}}_{j_1\dots j_{2n-1}}K^{j_1}_{i_1}\left[\frac{1}{2}R^{j_2 j_3}_{i_2i_3}- (t^2-1)K^{j_2}_{i_2}K^{j_3}_{i_3}\right]\cdots \left[\frac{1}{2}R^{j_{2n-2}j_{2n-1}}_{i_{2n-2}i_{2n-1}}- (t^2-1)K^{j_{2n-2}}_{i_{2n-2}}K^{j_{2n-1}}_{i_{2n-1}}\right]\, ,
\label{Qterm}
\end{equation}
where $R^{j_1 j_2}_{i_1i_2}$ is the projection of the $D$-dimensional curvature onto the boundary. Since we assume the curvature to be regular, the only divergences come now from the extrinsic curvatures. If we expand this expression we get
\begin{equation}
\mathcal{Q}_{n}=\sum_{l=1}^{n} \frac{c_l}{2l-1} \delta^{i_1\dots i_{2n-1}}_{j_1\dots j_{2n-1}}K^{j_1}_{i_1}\cdots K^{j_{2l-2}}_{i_{2l-2}}K^{j_{2l-1}}_{i_{2l-1}}R^{j_{2l} j_{2l+1}}_{i_{2l}i_{2l+1}}\cdots R^{j_{2n-2}j_{2n-1}}_{i_{2n-2}i_{2n-1}}\, ,
\label{Qterm}
\end{equation}
where the coefficients are
\begin{equation}\label{cl}
\begin{aligned}
c_l&=\frac{2n(2l-1)}{2^{n-l}}\binom{n-1}{l-1}\int_0^{1}dt(1-t^2)^{l-1}=\frac{1}{2^{n-3l+2}}\frac{n!(l-2)!}{(n-l)!(2l-1)!}
\end{aligned}
\end{equation}
Then, taking into account the previous observations we get the following result:
\begin{equation}
\begin{aligned}
&\lim_{r\rightarrow 0}\int_{\mathcal{B}_{\rm cap}}d^{d}x\sqrt{|h|}Q_{n}=\lim_{r\rightarrow 0}\int_{\mathcal{C}}d\sigma \int d\tilde\theta M Q_{n}\\
&=\int_{\mathcal{C}}d\sigma \int d\tilde\theta \sum_{l=1}^{n} c_l \delta^{\tilde\theta i_2 \dots i_{2n-1}}_{\tilde\theta j_2\dots j_{2n-1}}K^{j_2}_{i_2}\cdots K^{j_{2l-2}}_{i_{2l-2}}K^{j_{2l-1}}_{i_{2l-1}}R^{j_{2l} j_{2l+1}}_{i_{2l}i_{2l+1}}\cdots R^{j_{2n-2}j_{2n-1}}_{i_{2n-2}i_{2n-1}}\\
&=\int_{\mathcal{C}}d\sigma \int d\tilde\theta \sum_{l=1}^{n} c_l \delta^{\tilde\theta A_2 \dots A_{2n-1}}_{\tilde\theta B_2\dots B_{2n-1}}K^{B_2}_{A_2}\cdots K^{B_{2l-2}}_{A_{2l-2}}K^{B_{2l-1}}_{A_{2l-1}}R^{B_{2l} B_{2l+1}}_{A_{2l}A_{2l+1}}\cdots R^{B_{2n-2}B_{2n-1}}_{A_{2n-2}A_{2n-1}}\, ,
\end{aligned}
\end{equation}
where $d\sigma=d^{d-1}x\sqrt{|\sigma|}$ is the volume element on $\mathcal{C}$. Now, we may express the integrand using only intrinsic indices $A,B$ of the joint $\mathcal{C}$, so that $\delta^{\tilde\theta A_2 \dots A_{2n-1}}_{\tilde\theta B_2\dots B_{2n-1}}\rightarrow \delta^{A_2 \dots A_{2n-1}}_{B_2\dots B_{2n-1}}$. On the other hand, since $\tilde\theta$ is a local gaussian coordinate, the integral can only be performed within the local coordinate patch. However, we may just add up all the contributions from different patches in order to obtain the global integration along the cap, so that $\tilde\theta\rightarrow \theta$. Let us also introduce the schematic notation
\begin{equation}
K^{2l-2}R^{n-l}\equiv \delta^{A_2 \dots A_{2n-1}}_{B_2\dots B_{2n-1}}K^{B_2}_{A_2}\cdots K^{B_{2l-2}}_{A_{2l-2}}K^{B_{2l-1}}_{A_{2l-1}}R^{B_{2l} B_{2l+1}}_{A_{2l}A_{2l+1}}\cdots R^{B_{2n-2}B_{2n-1}}_{A_{2n-2}A_{2n-1}}\, .
\end{equation}
In this way, we can write 
\begin{equation}
\begin{aligned}
\lim_{r\rightarrow 0}\int_{\mathcal{B}_{\rm cap}}d^{d}x\sqrt{|h|}\mathcal{Q}_{n}=\int_{\mathcal{C}}d\sigma \int_{0}^{\Theta} d\theta \sum_{l=1}^{n} c_l K^{2l-2}R^{n-l}\, .
\end{aligned}
\end{equation}
Now, in the limit $r\rightarrow 0 $ the extrinsic curvatures are ill-defined since they depend  on the angle $\theta$. However, the integration can be actually performed by noting the following. The normal $n$ to the cap can be spanned by a linear combination of two different normals living in the $(r,\theta)$-plane: in particular we may use $n=a n_1+b s_1$. Since $n(\theta=0)=n_1$, we must have $a(0)=1$, $b(0)=0$. Also, the normal has unit norm $1=n^2=a^2+b^2$. In addition, the angle theta is defined by
\begin{equation}
\cos\theta=n(0)\cdot n(\theta)=a(0)a(\theta)+b(0)b(\theta)\, ,
\end{equation}
but $a(0)=1$ and $b(0)=0$, so that we conclude $a(\theta)=\cos\theta$, $b(\theta)=\pm \sin\theta$. Let us choose the $+$ sign, which corresponds to positive orientation, so that we can write the normal $n$ as
\begin{equation}
n=\cos\theta n_1+\sin\theta s_2\, .
\end{equation}
In this way, $n$ is interpolating between $n_1$ and $n_2$ when $\theta$ goes from $0$ to $\Theta$ --- see Figure \ref{Tmlj} (b). Also, this implies that the extrinsic curvature of $\mathcal{C}$ associated to $n$ can be decomposed in terms of those of $n_1$ and $s_1$:
\begin{equation}
K^{A}_{B}=\cos\theta L^{A}_{1\, B}+\sin\theta  Q^{A}_{1\, B}\, ,
\end{equation}
where 
\begin{equation}
L_{1\, AB}=e_{A}^{\mu}e_{B}^{\nu}\nabla_{\mu}n_{1\nu}\, ,\quad Q_{1\, AB}=e_{A}^{\mu}e_{B}^{\nu}\nabla_{\mu}s_{1\nu}\, ,
\end{equation}
and $e^A_{\mu}$ is a basis of tangent vectors of $\mathcal{C}$.
Since $L_1$ and $Q_1$ are two extrinsic curvatures of $\mathcal{C}$ associated to two orthogonal directions, the Gauss-Codazzi equations read
\begin{equation}\label{GC1}
R^{B_1B_2}_{A_1A_2}=\hat{\mathcal{R}}^{B_1B_2}_{A_1A_2}-2 L^{B_1}_{1\, [ A_1}L^{B_2}_{1\, A_2]}-2 Q^{B_1}_{1\, [A_1}Q^{B_2}_{1\, A_2]}\, ,
\end{equation}
where $\hat{\mathcal{R}}^{B_1B_2}_{A_1A_2}$ is the intrinsic curvature of $\mathcal{C}$. Now we are ready to compute the integral along the angle:
\begin{equation}
I_{\rm joint}=\int_{\mathcal{C}}d\sigma \int_{0}^{\Theta} d\theta \sum_{l=1}^{n}c_l \left(\cos\theta L_1+\sin\theta  Q_1\right)^{2l-2}R^{n-l}
\end{equation}
In order to proceed, it is convenient to write the trigonometric functions in terms of complex exponentials and then expand the product by using the binomial coefficients. We obtain a polynomial in powers of $e^{i\theta}$ which includes a constant term which is special as we are going to see. 
Then, the integration is straightforward and it yields
\begin{equation}
\begin{aligned}
I_{\rm joint}&=\int_{\mathcal{C}}d\sigma \Theta\sum_{l=1}^{n}c_l\binom{2l-2}{l-1}\frac{\left(L_1^2+Q_1^2\right)^{l-1}R^{n-l}}{2^{2l-2}}\\
&+\int_{\mathcal{C}}d\sigma \sum_{l=1}^{n}c_l{\rm Re}\left[\sum_{j=0}^{l-2}\binom{2l-2}{j}\frac{i(L_1-iQ_1)^j(L_1+iQ_1)^{2l-2-j}}{2^{2l-3}(2l-2-2j)}\left(e^{(2j-2l+2)i\Theta}-1\right)\right]R^{n-l}\, ,
\end{aligned}
\end{equation}
where in the first line we have the special term which is proportional to the total angle $\Theta$, while the other terms depend on trigonometric functions of $\Theta$. 
Now, by using the expression of the coefficients $c_l$ (\ref{cl}), we see that
\begin{equation}
c_l\binom{2l-2}{l-1}2^{-2l+2}=\frac{n}{2^{n-2}}\binom{n-1}{l-1}2^{l-1}\, ,
\end{equation}
and we can perform explicitly the summation appearing in the first line,
\begin{equation}
\sum_{l=1}^{n}\frac{n}{2^{n-2}}\binom{n-1}{l-1}2^{l-1}\left(L_1^2+Q_1^2\right)^{l-1}R^{n-l}=\frac{n}{2^{n-2}}\left(R+2L_1^2+2 Q_1^2\right)^{n-1}\, .
\end{equation}
 Then, according to (\ref{GC1}), we see that the combination between parenthesis is precisely the intrinsic curvature $\hat{\mathcal{R}}$ of $\mathcal{C}$. Therefore, this quantity is nothing but the $n-1$ Euler density of the induced metric on the joint $\mathcal{C}$
\begin{equation}
\hat{\mathcal{X}}_{2(n-1)}\equiv\frac{1}{2^{n-1}}\hat{\mathcal{R}}^{n-1}=\frac{1}{2^{n-1}}\delta^{A_1 \dots A_{2n-2}}_{B_1\dots B_{2n-2}}\mathcal{R}^{B_1B_2}_{A_1A_2}\cdots \mathcal{R}^{B_{2n-3}B_{2n-2}}_{A_{2n-3}A_{2n-2}}\, ,
\end{equation}
in terms of which we can write the result as
\begin{equation}
\begin{aligned}
I_{\rm joint}&=\int_{\mathcal{C}}d\sigma \Theta 2n\hat{\mathcal{X}}_{2(n-1)} \\
&+\int_{\mathcal{C}}d\sigma \sum_{l=1}^{n}c_l{\rm Re}\left[\sum_{j=0}^{l-2}\binom{2l-2}{j}\frac{i(L-iQ)^j(L+iQ)^{2l-2-j}}{2^{2l-3}(2l-2-2j)}\left(e^{(2j-2l+2)i\Theta}-1\right)\right]R^{n-l}\, .
\end{aligned}
\end{equation}
We still have to understand the role of the rest of the terms. For example, one may worry about the additivity of the action, since at first sight these terms do not seem to be additive. However, a closer inspection reveals that they actually are.\footnote{What we mean here by additivity is the property $I_{\rm joint}(\Theta_1+\Theta_2)=I_{\rm joint}(\Theta_1)+I_{\rm joint}(\Theta_2)$. However, due to the definition of the angle $\Theta$, the action is actually non-additive in the presence of timelike joints \cite{PhysRevD.50.4914}. The analogous property in the spacelike case (\ref{typeI}) ensures additivity of the action in the usual sense.}  We are expressing the result in terms of the extrinsic curvatures associated to the two normals adapted to the segment $\mathcal{B}_1$, but there is a more natural way to express it if we also make use of the orthonormal system associated to $\mathcal{B}_2$, $(n_2, s_2)$, related to $(n_1, s_1)$ according to (\ref{rot}).
The same relation will hold between the extrinsic curvatures,
\begin{equation}
L_2=L_1 \cos\Theta+Q_1\sin\Theta\, ,\quad Q_2=L_1 \sin \Theta-Q_1 \cos\Theta\, ,
\end{equation}
where $L_{2\, AB}=e_{A}^{\mu}e_{B}^{\nu}\nabla_{\mu}n_{2\nu}$, $Q_{2\, AB}=e_{A}^{\mu}e_{B}^{\nu}\nabla_{\mu}s_{2\nu}$. Then, let us note the following:
\begin{equation}
(L_1-iQ_1)^j(L_1+iQ_1)^{2l-2-j}e^{(2j-2l+2)i\Theta}=(L_2+iQ_2)^j(L_2-iQ_2)^{2l-2-j}\, .
\end{equation}
This allows us to write these terms in a more symmetric way:
\begin{equation}
\begin{aligned}
I_{\rm joint}&=\int_{\mathcal{C}}d\sigma \left[2n \Theta  \hat{\mathcal{X}}_{2(n-1)}+\mathcal{F}_n(L_1,Q_1)+\mathcal{F}_n(L_2,Q_2)\right]\, ,
\end{aligned}
\end{equation}
where, after some simplifications
\begin{equation}
\begin{aligned}
\mathcal{F}_n(L,Q)&=\sum_{l=2}^{n}\frac{n!(l-1)!}{2^{n-l-1}(n-l)!}R^{n-l}{\rm Im}\left[\sum_{j=0}^{l-2}\frac{(L-i Q)^j(L+i Q)^{2l-2-j}}{j!(2l-2-j)!(l-j-1)}\right]\, ,
\end{aligned}
\end{equation}
Therefore, these contributions do not actually depend on the angle of the joint, but they are related to the boundary of every segment. Expressed in this way, the contribution from the joints is explicitly additive. As we can see, the structure of the terms $\mathcal{F}_n$ is in general very complicated and it depends on both extrinsic and intrinsic curvatures. However, for Gauss-Bonnet gravity, this result takes a quite simple form:
\begin{equation}\label{jntGB}
\begin{aligned}
I_{\rm joint}&=4\int_{\mathcal{C}}d\sigma \left[\Theta\hat{\mathcal{R}}+2L^{[A}_{1\, A}Q^{B]}_{1\, B}+2L^{[A}_{2\, A}Q^{B]}_{2\, B}\right]\, .
\end{aligned}
\end{equation}
A remarkable property of the $n$-th Lovelock action is that, according to the Gauss-Bonnet-Chern theorem \cite{GBCTheorem}, in $D=2n$ dimensions it is the Euler characteristic of the manifold, the precise relation being
\begin{equation}
\mathcal{X}(\mathcal{M}_{2n})=\frac{1}{n! (4\pi)^n}I^{(n)}\left[\mathcal{M}_{2n}\right]\, .
\end{equation}
 If the manifold has boundaries, one needs to include the boundary term $\mathcal{Q}_n$ in order to really obtain the Euler characteristic \cite{10.2307/1990134}, and if further the boundary is nonsmooth one would need to include the terms that we have just derived. In appendix B we check that the Gauss-Bonnet action in $D=4$ with the joint terms (\ref{jntGB}) gives the right result for the Euler characteristic of a 4-dimensional cylinder deformed by an arbitrary function.


\subsection{Spacelike joints of type I}\label{23}
\begin{figure}[t]
  \centering
   \includegraphics[height=8cm]{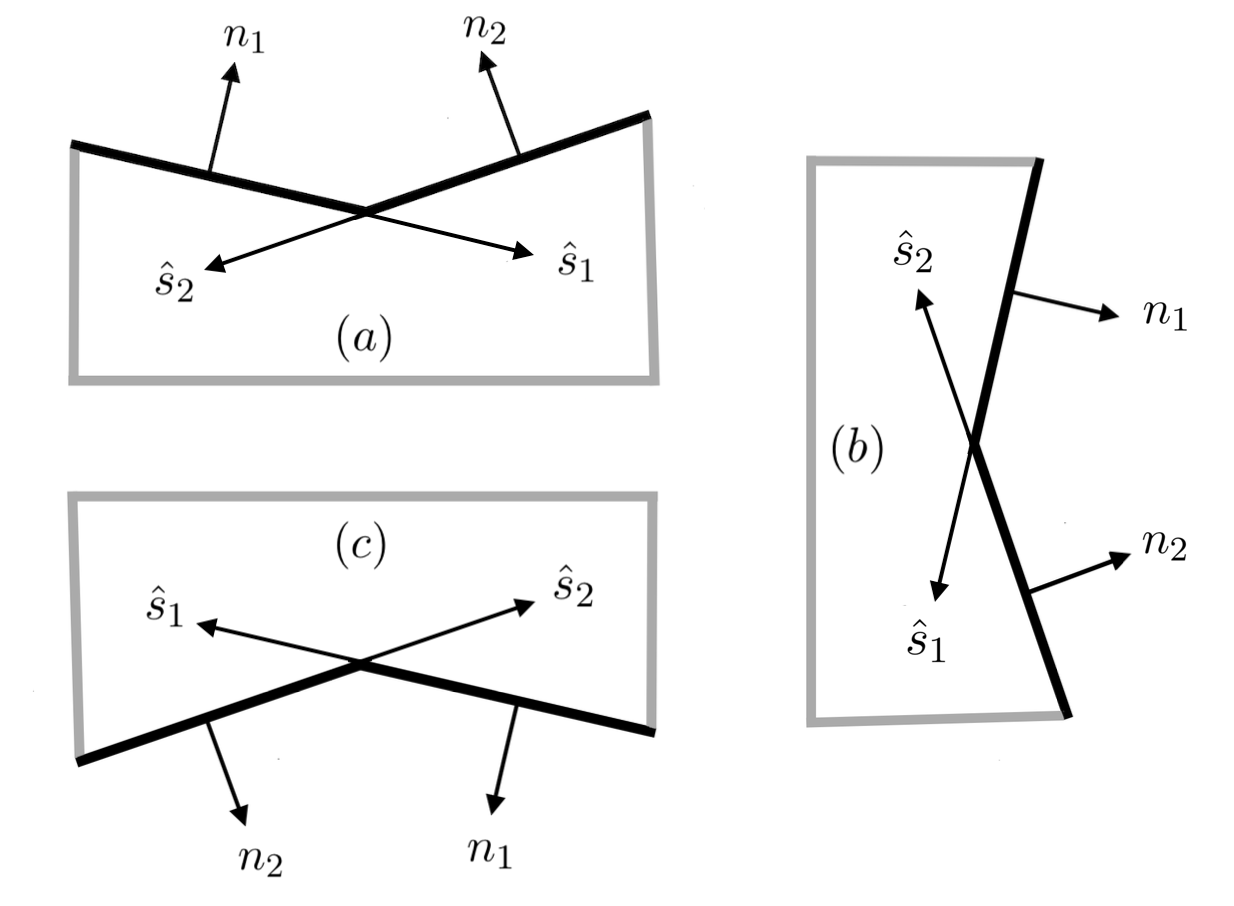}
  \caption{\small Spacelike joints of the type I. The vertical axis represents the timelike direction. We show the normal 1-forms $n=n_{\mu}dx^{\mu}$ and the tangent vectors $\hat s=s^{\mu}\partial_{\mu}$. In all the cases shown $\cosh(\eta)=|n_1\cdot n_2|$ and $\eta>0$.}
  \label{Splj}
\end{figure}

Let us consider now the case of spacelike joints as the ones illustrated in Figure \ref{Splj}. Here, either the normal  (cases (a) and (c)) or the tangent (case (b)) vectors are timelike. At the joint, the orthonormal frames adapted to the boundaries $\mathcal{B}_1$ and $\mathcal{B}_2$ are related by a Lorentz boost.  Let $(n_1, s_1)$ be the normal and tangent 1-forms in the first boundary that are normal to the joint, and let $(n_2, s_2)$ be those in the second boundary. We choose the normal 1-forms $n_{1,2}$ to point outside of the region of interest, and the tangent vectors $\hat s_{1,2}$ also point outside of the boundary at the joint. 
Then, at the joint, both basis are related by a boost of the form
\begin{equation}\label{boost}
\begin{aligned}
n_2&=n_1 \cosh\eta +s_1\epsilon' \sinh \eta \, ,\\
s_2&=- n_1 \epsilon' \sinh\eta -s_1 \cosh \eta \, ,
\end{aligned}
\end{equation}
with certain rapidity parameter $\eta$ and where $\epsilon'=\pm1$ is a sign so that when we increase $\eta$ the basis rotates with positive orientation. In order to determine it, let us consider 2-dimensional Minkowski space with metric $ds^2=-dt^2+dx^2$ and volume element $dt\wedge dx$. The change of variables $t=\tau \cosh\eta$, $x=\tau \sinh\eta$ has positive orientation and for fixed $\tau>0$ it describes a surface whose normal is $n=dt \cosh\eta-dx \sinh\eta$. This is the situation considered in Figure \ref{Splj} (a) identifying $n=n_2$, $dt=n_1$, $dx=s_1$. Thus, we get $\epsilon'=-1$. In the case (c) the result is the same, since the difference is an overall sign in the change of coordinates. When the normal is spacelike, the appropriate parametrization is instead $t=-\rho \sinh\eta$, $x=\rho \cosh\eta$ and the surfaces of constant $\rho>0$ have a normal 1-form $n=\cosh\eta dx+\sinh \eta dt$. Thus, in Figure \ref{Splj} (b) we identify $n=n_2$, $dx=n_1$ and $dt=s_1$ and the sign is $\epsilon'=+1$ instead. The conclusion is that we have to choose the sign equal to the signature of the normal $\epsilon'=n^2\equiv \epsilon$.

Then, the idea is again to replace this sharp corner by a smooth cap, in which the orthonormal frame interpolates from $(n_1, s_1)$ to $(n_2, s_2)$ and at the end take the limit in which the cap reduces to a point. The metric of such smoothed boundary can be written locally as
\begin{equation}
ds^2=\epsilon N^2d\rho^2+h_{ij}dx^idx^j\, ,\quad h_{ij}dx^idx^j=-\epsilon M^2 d\tilde\eta^2+\sigma_{AB}dx^{A}dx^{B}\, ,
\end{equation}
where the normal vector is $n=N d\rho$, whose signature is $\epsilon=+1$ for timelike boundaries and $\epsilon=-1$ for spacelike ones.\footnote{In the case of spacelike boundaries $n=Nd\rho$ is future directed, so according to our conventions this formula is valid when the region of the space time is in the past of the boundary. When the boundary is in the past instead, we should choose $n_{\mu}$ to be past directed, which introduces an additional sign in the extrinsic curvature. However, another sign is introduced as a consequence of the change of orientation when we choose $n=-N d\rho$. Then, the result (\ref{Fnsp}) has the same form when it is expressed in terms of the outwards-directed 1-form $n_{\mu}$. See also \cite{Jee1984,10.2307/44237512,birman1984,Lehner:2016vdi} for a more detailed discussion on these sign issues.} Regularity at $\rho=0$ imposes similar conditions as before, in particular\footnote{Actually, we only need that $\lim_{\rho\rightarrow 0}\frac{\partial_{\rho} M}{N}=c>0$, but if we want to identify $\tilde\eta$ with the boost parameter $\eta$ we have to choose the normalization $c=1$. } 
\begin{equation}
\lim_{\rho\rightarrow 0}\frac{\partial_{\rho} M}{N}=1\, , \quad M\Big|_{\rho\rightarrow 0} =M(\rho)\, , \quad M\Big|_{\rho= 0} =0\,
\end{equation}
The extrinsic curvature is
\begin{equation}
K_{ij}=\frac{\epsilon}{2N}\partial_{\rho} h_{ij}\, .
\end{equation}
and the component $K^{\tilde \eta}_{\tilde \eta}=\epsilon\frac{\partial_{\rho} M}{M N}\rightarrow \frac{\epsilon}{M}$ diverges when $\rho\rightarrow 0$. The computation is very similar to the one performed in the previous subsection and we get in this case
\begin{equation}
I_{\rm joint}=\int_{\mathcal{C}}d\sigma \int_{0}^{\eta} d\eta' \sum_{l=1}^{n}c_l \epsilon^{l} \left(\cosh \eta' L_1+\sinh \eta'  \epsilon Q_1\right)^{2l-2}R^{n-l}\, ,
\end{equation}
where, as before, $L_{1\, AB}=e_{A}^{\mu}e_{B}^{\nu}\nabla_{\mu}n_{1\nu}$ and $Q_{1\, AB}=e_{A}^{\mu}e_{B}^{\nu}\nabla_{\mu}s_{1\nu}$ are the extrinsic curvatures associated to the normal and to the tangent in the first segment. In order to perform the integration, we express the hyperbolic trigonometric functions in terms of exponentials and expand the product. The result reads
\begin{equation}
\begin{aligned}
I_{\rm joint}&=\int_{\mathcal{C}}d\sigma 2n\epsilon\eta \hat{\mathcal{X}}_{2(n-1)} \\
&-\int_{\mathcal{C}}d\sigma \sum_{l=1}^{n}\frac{n!(l-1)! R^{n-l}}{(n-l)!2^{n-l}}\sum_{0=j\neq l-1}^{2l-2}\frac{(L_1+\epsilon Q_1)^j(L_1-\epsilon Q_1)^{2l-2-j}\epsilon^{l}}{j!(2l-2-j)!(l-1-j)}\left[e^{-2(l-1-j)\eta}-1\right]
\end{aligned}
\end{equation}
where we used the Gauss-Codazzi equation
\begin{equation}
R^{B_1B_2}_{A_1A_2}=\mathcal{R}^{B_1B_2}_{A_1A_2}-2\epsilon L^{B_1}_{1\, [A_1}L^{B_2}_{1\, A_2]}+2 \epsilon Q^{B_1}_{1\, [A_1}Q^{B_2}_{1\, A_2]}\, 
\end{equation}
in order to write the result in terms of the Euler density $\hat{\mathcal{X}}_{2(n-1)}$ of the induced metric. Finally, if we take into account the relation between the orthogonal systems (\ref{boost}),
\begin{equation}
\begin{aligned}
L_2&=L_1 \cosh\eta +Q_1\epsilon \sinh \eta \, ,\\
Q_2&=- L_1 \epsilon \sinh\eta -Q_1 \cosh \eta \, ,
\end{aligned}
\end{equation}
we can write the contribution from the joint as
\begin{equation}\label{typeI}
\begin{aligned}
I_{\rm joint}&=\int_{\mathcal{C}}d\sigma \left[2n\epsilon \eta \hat{\mathcal{X}}_{2(n-1)}+\mathcal{F}_{n}(L_1,Q_1)+\mathcal{F}_{n}(L_2, Q_2)\right]\, ,
\end{aligned}
\end{equation}
where
\begin{equation}\label{Fnsp}
\begin{aligned}
\mathcal{F}_n(L, Q)&= \sum_{l=2}^{n}\frac{n!(l-1)! R^{n-l}\epsilon^{l-1}}{(n-l)!2^{n-l}}\sum_{0=j\neq l-1}^{2l-2}\frac{(L+ Q)^j(L-Q)^{2l-2-j}}{j!(2l-2-j)!(l-1-j)}\, .\end{aligned}
\end{equation}
Had we considered a spacelike boundary segment placed at the past of the region of interest --- see Figure \ref{Splj} (c) ---, the result would have been the same, as long as the 1-form normals $n_{1,2}$ point outside of the region (this is, to the past). 

 \subsection{General case}\label{24}
In the previous two sections we have computed the contribution from timelike and type I spacelike joints by using the smoothing method employed by Hayward \cite{Hayward:1993my}, but it would be important to check that these terms actually make the variational problem well-posed. Fortunately, it was shown in \cite{Ruan:2017tkr} that the two methods are equivalent, so we can rely on the joint terms we have found.
The problem is that the smoothing method is not directly applicable to type II joints, \textit{e.g.} where the normal goes from spacelike to timelike, since it is not possible to describe a boundary interpolating smoothly in that case. We would be forced to examine the variational problem in the presence of corners and identify which terms must be added. However, with the results we have accumulated so far it is possible to derive the general form of the contribution for all types of joints without making an explicit use of the variational method. As we have seen, for type I spacelike joints, the contribution to the action has the form

\begin{equation}\label{Ijoint1}
\begin{aligned}
I_{\rm joint}&=\int_{\mathcal{C}}d\sigma \left[2n\eta \hat{\mathcal{X}}_{2(n-1)}+\mathcal{F}_{n, 1}+\mathcal{F}_{n,2}\right]\, ,
\end{aligned}
\end{equation}
where $\eta=\pm \operatorname{arccosh}|n_1\cdot n_2|$ and $\mathcal{F}_{n, 1,2}$ are certain quantities which depend on the intrinsic and extrinsic curvatures of the joint adapted to the frames of each side of the joint. Let us note that these terms do not really depend on the change in the normal, but they are associated to the boundary of each segment and they actually only depend on the kind of segment. In this sense, it seems natural to re-arrange the boundary and joint terms in the action as

\begin{equation}\label{brjnt}
I_{\rm bdry}=\sum_{k}\left[\int_{\mathcal{B}_k}d\Sigma \mathcal{Q}_{n}+\int_{\partial\mathcal{B}_k}d\sigma \mathcal{F}_{n}\right]\, ,\quad I_{\rm joint}=\sum_{l}\int_{\mathcal{C}_l}d\sigma 2n\eta \hat{\mathcal{X}}_{2(n-1)}\, ,
\end{equation}
so that every segment contains also a contribution coming from its boundary. Of course, this is equivalent to (\ref{Ijoint1}) since the joints are intersections of two segments, each one contributing with its own $\mathcal{F}_n$. Since $\mathcal{F}_{n}$ only depends on geometric quantities defined on $\mathcal{B}_k$ and its boundary, it is clear that when we consider variations of the action these terms are independent for every segment $\mathcal{B}_k$. We can interpret this fact from the point of view variational principle if we recall the total derivative $D_i \dslash H^i$ in (\ref{var}). For every segment of boundary $\mathcal{B}_k$, this term can be integrated by using Stokes' theorem, yielding the integral of $s_i \dslash H^i$ over  $\partial\mathcal{B}_k$.  Then, it seems clear that some part of this term can be arranged as a total variation on its own, and this will give rise to $\mathcal{F}_n$. The other part of this term will produce a total variation only when it is combined with the contribution coming from the other boundary at the joint, and the resulting term is then $2n\eta \hat{\mathcal{X}}_{2(n-1)}$. The conclusion of this observation is that the expressions for $\mathcal{F}_n$ are really independent of the kind of joint, and therefore the results that we have obtained are actually valid in general.

Then, we only need to determine the correct generalization of the term $2n\eta \hat{\mathcal{X}}_{2(n-1)}$ in (\ref{brjnt}) for type II joints. It seems natural that the form of this term will be actually the same, with the parameter $\eta$ being the same one that appears in the Einstein gravity case \cite{Lehner:2016vdi}. We can actually prove this point by taking into account the following observation. In $D=2n$ dimensions, the Lovelock action  $I^{(n)}[\mathcal{M}_{2n}]$ is, up to a constant factor, the Euler characteristic of the manifold $\mathcal{M}_{2n}$. A property of Euler's characteristic is its factorization rule for product spaces: we have $\mathcal{X}(\mathcal{A}\times\mathcal{B})=\mathcal{X}(\mathcal{A})\mathcal{X}(\mathcal{B})$. At the level of the Lovelock action, this means that in $D=2n$ dimensions it should factorize as $I^{(n)}[\mathcal{M}_{2k}\times\mathcal{M}_{2n-2k}]\propto I^{(k)}[\mathcal{M}_{2k}]I^{(n-k)}[\mathcal{M}_{2n-2k}]$. Since the form of the boundary and joint terms is the same in any dimension, we can use this fact to relate them between different Lovelock gravities --- in particular, we can relate them to the Einstein gravity ones. Therefore, let us consider the complete Lovelock action 
\begin{equation}
I^{(n)}=\int_{\mathcal{M}_{2n}}d^{2n}x\sqrt{|g|}\mathcal{X}_{2n}+ \sum_k\left[\int_{\mathcal{B}_k}d^{2n-1}x\sqrt{|h|}\mathcal{Q}_{n}+\int_{\partial \mathcal{B}_k}d\sigma \mathcal{F}_n\right]+\sum_l\int_{\mathcal{C}_l}d\sigma 2n\psi_l \hat{\mathcal{X}}_{2(n-1)}\, ,
\end{equation}
where we are assuming that the joint term contains the Euler density $\hat{\mathcal{X}}_{2(n-1)}$ weighted by some unknown quantities $\psi_l$. Then, let us evaluate $I^{(n)}$ on a product manifold $\mathcal{M}_{2n-2}\times\mathcal{M}_2$, where the first factor is assumed to be compact while the second one is allowed to have a nonsmooth boundary. The Euler density decomposes as $\mathcal{X}_{2n}=n \hat R\hat{\mathcal{X}}_{2(n-1)}$. On the other hand, the extrinsic curvature on the boundary has rank 1, and therefore, only the term in $ \mathcal{Q}_{n}$ containing one extrinsic curvature survives. Thus, we obtain $ \mathcal{Q}_{n}=2n \hat K \hat{\mathcal{X}}_{2(n-1)}$. Finally, the extrinsic curvatures of the joints vanish and therefore $\mathcal{F}_n=0$ and we get
\begin{equation}
I^{(n)}=n\int_{\mathcal{M}_{2n-2}}d^{2n-2}x\sqrt{|g|}\hat{\mathcal{X}}_{2(n-1)}\Bigg[\int_{\mathcal{M}_{2}}d^{2}x\sqrt{|g|}\hat{R} +2\int_{\partial \mathcal{M}_2}ds \hat K+2\sum_l\psi_l\Bigg]\, .
\end{equation}
hence the reduction of the $2n$-dimensional Lovelock action gives us the 2-dimensional Einstein action,
\begin{equation}
I^{(2)}=\int_{\mathcal{M}_{2}}d^{2}x\sqrt{|g|}\hat{R} +2\int_{\partial \mathcal{M}_2}ds  \hat K+2\sum_l\psi_l\, .
\end{equation}
We obtain the York-Gibbons-Hawking boundary term and the contributions $\psi_l$ coming from the corners. Then, consistency demands that the quantities $\psi_l$ in the Lovelock joint term are the same as in Einstein gravity. Since these contributions are already known for Einstein gravity for all kind of joints \cite{Lehner:2016vdi}, we have determined the action in the most general case for spacelike and timelike boundaries. The result is actually stronger, since it can be applied as well to null boundaries. In that case, it tells us that the value of $\psi$ for intersections of a null boundary and any other kind of boundary is $\psi\equiv a\propto \log|k\cdot n|$, where $k$ and $n$ are the respective normals. For more details about how to find this parameter in all cases, we refer to the original work \cite{Lehner:2016vdi}.
Putting it all together, we obtain the result for the Lovelock action explained in \ref{CLove}.

Let us mention that the method of reduction of the action can also be used to to obtain some information about the boundary terms in the null case. We learn that, when evaluated on the product geometry $\mathcal{M}_{2n-2}\times\mathcal{M}_2$, the surface term in the null case should reduce to
\begin{equation}
I_{\rm bdry}=-\int d\lambda d^{d-1}x\sqrt{\sigma} 2n\kappa \hat{\mathcal{X}}_{2(n-1)}\, ,
\end{equation}
where $\kappa$ is defined by the relation $k^{\alpha}\nabla_{\alpha}k_{\beta}=\kappa k_{\beta}$. However, this result does not characterize completely the null boundary term.

\section{Complexity of global AdS}\label{3}

As we remarked in the introduction, one of the situations where one needs to compute the gravitational action in a nonsmooth region appears in the context of holographic complexity.
Complexity is a quantity that has its origin in quantum information science, but in recent years a growing interest has been focused in extending this concept to QFTs, and in particular to holographic CFTs. The main property that has been noted is the correspondence between the growth of the AdS-Einstein-Rosen bridge at late times and the growth of complexity in a quantum theory. This has motivated two different proposals for holographic duals of the complexity in the frame of the AdS/CFT correspondence. The ``complexity=volume'' proposal \cite{Susskind:2014rva,Stanford:2014jda,Susskind:2014jwa} states the equivalence between complexity in the CFT and certain extremal volume in the bulk, while according to the ``complexity = action'' proposal \cite{Brown:2015bva,Brown:2015lvg} the complexity is given by the gravitational action in the Wheeler-DeWitt patch of AdS --- see the references above and \textit{e.g.} \cite{Chapman:2016hwi,Carmi:2016wjl} for more details. The latter has the advantage of being also applicable to the case of higher-curvature gravity, while the former will probably need of a non-trivial modification if we decide to include these corrections in the bulk--- see \cite{Bueno:2016gnv} for a possible generalization. 
Then, let us focus on the ``complexity = action'' proposal. More precisely, it states that the complexity in a fixed time slice $\Sigma$ of a holographic CFT is equal to
\begin{equation}
\mathcal{C}_A(\Sigma)=\frac{I_{\rm WDW}}{\pi \hbar}\, ,
\end{equation}
where $I_{\rm WDW}$ is the gravitational action in the Wheeler-DeWitt patch: the causal diamond attached to the constant time slice $\Sigma$ in the boundary of AdS. The structure of the divergences of the complexity was studied in \cite{Carmi:2016wjl}, where it was found that there is a double series expansion due to the presence of a logarithmic term:
\begin{equation}\label{Cexp}
\mathcal{C}_A\sim\frac{a_1}{\delta^{d-1}}+\frac{a_2}{\delta^{d-3}}+\ldots+\log\left(\frac{\ell}{\delta}\right)\left[\frac{b_1}{\delta^{d-1}}+\frac{b_2}{\delta^{d-3}}+\ldots\right]\, ,
\end{equation}
where $\ell$ is an arbitrary length scale, associated to the freedom to choose the normalization of the null boundary generators. In general the coefficients $a_i$, $b_i$ involve integrals in $\Sigma$ of intrinsic and extrinsic geometric quantities but particularizing to global AdS they are just numerical factors. It was observed that the coefficients $a_i$ in the first series are actually dependent on the regularization scheme and they can also be modified if we for example rescale the length $\ell$ in the logarithmic expansion. On the other hand, the expansion containing the logarithm seems to be regulator-independent. Therefore, it would be interesting to search for universal terms in this series, this is, terms which are unaffected by rescalings of $\delta$.  A prototypical example is the case of holographic entanglement entropy. When studying the structure of the divergences of the EE across an $\mathbb{S}^{d-1}$ one finds that there is a universal piece\footnote{See \cite{Casini:2015woa} for a detailed discussion about the meaning of universal terms.} that reads \cite{Ryu:2006ef,Nishioka:2009un,Myers:2010tj,Myers:2010xs,CHM}
\begin{equation}\label{asta}
S_{\rm univ}=\begin{cases}
(-1)^{\frac{d-2}{2}} 4a^*_d \log(R/\delta) \quad &\text{for even } d \, , \\
 (-1)^{\frac{d-1}{2}}2\pi a^*_d \quad &\text{for odd } d\, .
\end{cases}
\end{equation}
The constant $a^*_d$ can be read from the gravitational Lagrangian according to the relation 
\begin{equation}\label{astar}
a^*_d=-\frac{\pi^{d/2}\tilde{L}^{d+1}}{d \Gamma(d/2)}\mathcal{L}
|_{{\rm AdS}}\, ,
\end{equation}
valid at least for even $d$ \cite{Imbimbo:1999bj,Schwimmer:2008yh}, and probably also for odd $d$ \cite{Myers:2010tj,Myers:2010xs,Bueno:2018xqc}. In the case of Einstein gravity holography, this constant is just proportional to $\tilde L^{d-1}/G$, and so is any other ``central charge''. The introduction of higher-curvature terms with arbitrary couplings breaks this degeneracy and allows to search for relations between CFT quantities. For example, $a^*_d$ coincides with the $a$-type trace-anomaly charge in even dimensional theories  \cite{Myers:2010tj,Myers:2010xs}.
Hence, it would be interesting to explore if there are any other universal constants playing a similar role in the case of holographic complexity, and what kind of information about the underlying CFT they contain. In order to do so, let us consider Lovelock gravity with action
\begin{equation}
I_{\rm bulk}=\frac{1}{16\pi G}\int_{\mathcal{M}}d^{d+1}x\sqrt{|g|}\left\{\frac{d(d-1)}{L^2}+R+\sum_{n=2}^{\lfloor d/2\rfloor}\frac{\lambda_n L^{2n-2}(-1)^{n}}{(d-2)\ldots(d+1-2n)}\mathcal{X}_{2n}\right\}\, .
\end{equation}
Let us denote by $\tilde L$ the length scale of the AdS vacua of this theory, which is determined by $L$ and by the couplings according to $\tilde L^2=L^2/f_{\infty}$, where $f_{\infty}$ is the solution of the equation
\begin{equation}
1-f_{\infty}+\sum_{n=2}^{\lfloor d/2\rfloor}\lambda_ nf_{\infty}^n=0\, 
\end{equation}
such that $f_{\infty}\rightarrow 1$ when $\lambda_n\rightarrow 0$.
Then, the metric of global AdS is
\begin{equation}\label{GAdS}
ds^2=\frac{\tilde L^2}{\cos^2\theta}\left(-d\tau^2+d\theta^2+\sin^2\theta d\Omega_{(d-1)}^2\right)\, ,
\end{equation}
with $\theta\in[0,\pi/2)$, $\tau\in[-\pi/2,\pi/2)$.
We would like to evaluate the gravitational action on a regularized Wheeler-DeWitt patch, so that there is a cutoff distance $\delta$ from the boundary. There are several ways to regularize the WDW patch, which gives rise to some ambiguities in the complexity. We can consider the same regularizations as in \cite{Carmi:2016wjl}
\begin{eqnarray}
\mathcal{W}_1(\delta')&=&\left\{(\tau, \theta, \phi_i)\in \text{AdS}_{d+1}/\, \, \, |\tau|\le \pi/2-\theta,\, \, 0\le\theta\le \pi/2-\delta'\right\}\, ,\\
\mathcal{W}_2(\delta')&=&\left\{(\tau, \theta, \phi_i)\in \text{AdS}_{d+1}/ \, \, \, |\tau|\le \pi/2-\delta'-\theta,\, \, 0\le\theta\le \pi/2-\delta'\right\}
\end{eqnarray}
In the first case, the boundary consists of three pieces: the null boundaries $S_{\pm}$ given by $\tau=\pm(\pi/2-\theta)$, $\theta\in(0,\pi/2-\delta')$ and the timelike boundary $R: \theta=\pi/2-\delta'$, $\tau\in(-\delta',\delta')$. There are joints at $\theta=\pi/2-\delta'$, $\tau=\pm \delta'$. In the second region we only have the null boundaries $S'_{\pm}: \tau=\pm(\pi/2-\delta'-\theta)$ and their joint at $\tau=0$, $\theta=\pi/2-\delta'$.\footnote{The boundary also contains two caustics at the tips of the WDW patch where all of the null generators meet. By performing a computation similar to the one in \cite{Chapman:2016hwi} it can be shown that the contribution from these caustics coming from the Euler density $\mathcal{X}_{2n}$ is a constant in the critical dimension $D=2n$ and that it vanishes for $D>2n$. We do not include the Euler density in the critical dimension, so we can safely ignore these caustics.} The complete action then has three contributions $I=I_{\rm bulk}+I_{\rm bdry}+I_{\rm joint}$. While we already can determine $I_{\rm bulk}$ and $I_{\rm joint}$, the problem is of course that we do not know the surface terms for null boundaries.\footnote{Here we consider the terms $\mathcal{F}_n$ as part of the surface term, as discussed previously.} In the case of Einstein gravity, the null surface terms vanish as long as we choose the null generators to be affinely parametrized. However, it is unclear that the same behavior will happen in Lovelock gravity. We will remain agnostic about the null boundary terms, but a great deal of information can be obtained already from the bulk and joint contributions. In particular, as observed in \cite{Carmi:2016wjl}, the logarithmic part of the series (\ref{Cexp}) comes only from $I_{\rm joint}$.\footnote{There is also one logarithmic term coming from the bulk action when $d$ is odd --- see (\ref{logbulk}).} On general ground, we can expect the same thing to happen here, \textit{i.e.}, we do not expect that the (unknown) boundary term contributes to the logarithmic series, which means that we can already determine it by using our results. The contribution from the joints reads
\begin{equation}
I_{\rm joint}=\frac{1}{8\pi G}\int_{\mathcal{C}}d\sigma a \left[1+\sum_{n=2}^{\lfloor d/2\rfloor}\frac{n \lambda_n L^{2n-2}(-1)^{n}}{(d-2)\ldots(d+1-2n)}\hat{\mathcal{X}}_{2(n-1)}\right]\, ,
\end{equation}
where $a$ is determined in the same way as in \cite{Lehner:2016vdi,Carmi:2016wjl}. In particular, in the joint between two null boundaries we have $a=\pm\log|k_1\cdot k_2/2|$, where $k_{1,2}$ are the outwards-directed null 1-forms normal to the boundaries and the sign depends on the case. In the joint between a null boundary and a timelike surface with normal $n$ we have instead $a=\pm\log|n\cdot k_1|$. Then, let us parametrize the null normals as $k_1=\alpha_1\tilde L(d\theta+d\tau)$, $k_2=\alpha_2\tilde L (d\theta-d\tau)$, and we also have, in the regularization $\mathcal{W}_1$: $n=\frac{\tilde L}{\sin\delta'}d\theta$. Therefore, in the first regularization we obtain two joints whose parameters $a$ read $a_{1,2}=-\log|n\cdot k_{1,2}|=-\log|\alpha_{1,2} \sin\delta'|$, so that when we add them up we get $a_1+a_2=2\log\left(\frac{1}{\sqrt{\alpha_1\alpha_2}\sin\delta'}\right)$. The same result is obtained in the other regularization, where we only have one joint, but its contribution is $a=-\log|k_1\cdot k_2/2|=a_1+a_2$. Then, the corner contribution is independent from the scheme. The induced metric on $\mathcal{C}$ is
\begin{equation}
^{(d-1)}ds^2=\tilde L^2 \cot^2\delta' d\Omega_{(d-1)}^2\, ,
\end{equation}
so it is an sphere of radius $\tilde L \cot\delta'$, which implies that the reduced Euler densities take the values $\hat{\mathcal{X}}_{2(n-1)}=\frac{\tan^{2(n-1)}(\delta')}{\tilde L^{2(n-1)}}(d-1)\ldots(d+2-2n)$. Hence, the contribution from the joint is
\begin{equation}
I_{\rm joint}=\frac{\tilde L^{d-1}\Omega_{(d-1)}}{4\pi G} \log\left(\frac{1}{\sqrt{\alpha_1\alpha_2}\sin\delta'}\right)\cot^{d-1}\delta'\left[1+\sum_{n=2}^{\lfloor d/2\rfloor}\frac{(d-1)n \lambda_n f_{\infty}^{n-1}(-1)^{n}}{(d+1-2n)}\tan^{2(n-1)}(\delta')\right]\, ,
\end{equation}
where $\Omega_{(d-1)}=2\pi^{d/2}/\Gamma(d/2)$ is the volume of $\mathbb{S}^{d-1}$.
Now, as in \cite{Carmi:2016wjl}, let us make the change of variables $z=\frac{2\tilde L\cos\theta}{1+\sin\theta}$, which takes the metric (\ref{GAdS}) to a standard Fefferman-Graham form 
\begin{equation}\label{FG}
ds^2=\frac{\tilde L^2}{z^2}\left[dz^2-\left(1+\frac{z^2}{4\tilde L^2}\right)^2dt^2+ \tilde L^2\left(1-\frac{z^2}{4\tilde L^2}\right)^2d\Omega_{(d-1)}^2\right]\, ,
\end{equation}
where we also introduced $t=\tilde L \tau$. Then, we rewrite the short-distance regulator $\delta'$ of global AdS in terms of the regulator for the previous metric $\delta=\frac{2\tilde L\sin\delta'}{1+\cos\delta'}$, which implies
\begin{equation}
\cot\delta'=\frac{\tilde L}{\delta}-\frac{\delta}{4\tilde L}\, ,\quad \csc\delta'=\frac{\tilde L}{\delta}+\frac{\delta}{4\tilde L}\, .
\end{equation}
In this way, we can write the exact result as
\begin{equation}
I_{\rm joint}=\frac{\tilde L^{d-1}\Omega_{(d-1)}}{4\pi G} \log\left(\frac{\tilde L+\delta^2/(4\tilde L)}{\sqrt{\alpha_1\alpha_2}\delta}\right)\sum_{n=1}^{\lfloor d/2\rfloor}\frac{(d-1)n \lambda_n f_{\infty}^{n-1}(-1)^{n}}{(d+1-2n)}\left(\frac{\tilde L}{\delta}-\frac{\delta}{4\tilde L}\right)^{d+1-2n} \, ,
\end{equation}
where we have introduced the $+1$ in the sum, with the convention that $\lambda_1=-1$. Now we expand this expression when $\delta\rightarrow 0$ and we obtain the following logarithmic series in the complexity
\begin{equation}
\begin{aligned}
\mathcal{C}_A&=\frac{\tilde L^{d-1}\Omega_{(d-1)}}{4\pi^2 G}\log\left(\frac{\tilde L}{\sqrt{\alpha_1\alpha_2}\delta}\right)\Bigg[\frac{\tilde L^{d-1}}{\delta^{d-1}}-\frac{(d-1)}{4}\frac{\tilde L^{d-3}}{\delta^{d-3}}\left(1-\frac{8\lambda_2 f_{\infty}}{d-3}\right)\\
&+\frac{(d-1)(d-2)}{32}\frac{\tilde L^{d-5}}{\delta^{d-5}}\left(1-\frac{16\lambda_2 f_{\infty}}{d-2}-\frac{96 \lambda_3f_{\infty}^2}{(d-2)(d-5)}\right)+\ldots\Bigg]+[\ldots]\, ,
\end{aligned}
\end{equation}
where $[\ldots]$ denotes the non-logarithmic $1/\delta$ series expansion. Note that the higher-order Lovelock terms do not modify the leading order Einstein gravity behavior. The reason is that the joint term involves intrinsic curvatures, which are vanishing for $\delta\rightarrow 0$. However, the subleading terms are modified: the $n$-th Lovelock density starts appearing in the coefficient of $1/\delta^{d+1-2n}$. In particular, when $d$ is odd the previous series contains a constant term that possibly will be universal, and all Lovelock densities contribute to this term. The universal term is in general given by
\begin{equation}\label{CAuniv1}
\mathcal{C}_A^{\rm univ}=(-1)^{\frac{d-1}{2}}b_d \log\left(\frac{\tilde L}{\sqrt{\alpha_1\alpha_2}\delta}\right)+[\ldots]\, ,
\end{equation}
where
\begin{equation}\label{Cd}
b_d=-\frac{\tilde L^{d-1}\Omega_{(d-1)}}{4\pi^2 G}\sum_{n=1}^{(d-1)/2}\lambda_n f_{\infty}^{n-1}\frac{n(d-1)(d-2n)!}{((d+1)/2-n)!^22^{d+1-2n}}\, ,
\end{equation}
and we leave open the possibility of having additional terms in $[\ldots]$ coming from the bulk or boundary contributions. Some of the first values of this constant are $b_d\times \frac{4\pi^2 G}{\tilde L^{d-1}\Omega_{(d-1)}}=\frac{1}{2}\, ,\frac{3}{8}-2\lambda_2 f_{\infty}\, ,\frac{5}{16}-\frac{9 \lambda _2 f_{\infty}}{8}-\frac{9 \lambda _3 f_{\infty}^2}{2}$ for $d=3$, $5$ and $7$ respectively. We may compare this constant to some known central charges, such as $a_d^*$ (\ref{astar}), the constant $C_T$ controlling the stress-energy tensor two-point function or the thermal entropy charge $C_S$, defined by the relation between entropy and temperature in a thermal plasma $s=C_S T^{d-1}$. For a holographic CFT dual to Lovelock gravity, these charges read
\begin{eqnarray}
a^*_d&=&-\frac{\pi^{(d-2)/2}\tilde L^{d-1}}{8\Gamma(d/2) G}\sum_{n=1}^{\lfloor d/2\rfloor}\lambda_n f_{\infty}^{n-1}\frac{(d-1)n}{(d+1-2n)}\, ,\\
C_T&=&-\frac{\Gamma(d+2)\tilde L^{d-1}}{8(d-1)\Gamma(d/2)\pi^{(d+2)/2 }G}\sum_{n=1}^{\lfloor d/2\rfloor}n\lambda_n f_{\infty}^{n-1}\, \\
C_S&=&\frac{\Gamma(d+1)\pi^{(2d-1)/2}2^{d-3}}{\Gamma\left(\frac{d+1}{2}\right)\Gamma(d/2)d^d}\frac{f_{\infty}^{d-1}\tilde L^{d-1}}{G}\, ,
\end{eqnarray}
where we recall that $\lambda_1=-1$ corresponds to the Einstein gravity contribution. We see that there is no simple way to express $b_d$ as a combination of these charges --- the dependence on the curvature order $n$ is very different --- so we conclude that this constant could be a new central charge characterizing the CFT.

For the sake of completeness we may also search for universal terms coming from the bulk part of the action, whose general form is
\begin{equation}
I_{\rm bulk}=
\begin{cases}
-4da^*_d\int_{\delta'}^{\pi/2}d\theta'\theta'\frac{\cos^{d-1}\theta'}{\sin^{d+1}\theta'}\quad &\text{if}\,\, \mathcal{W}=\mathcal{W}_1(\delta')\\
-4da^*_d\int_{\delta'}^{\pi/2}d\theta'\theta'\frac{\cos^{d-1}\theta'}{\sin^{d+1}\theta'}+4a^*_d \delta' \cot^d\delta' \quad &\text{if}\,\, \mathcal{W}=\mathcal{W}_2(\delta')\, ,
\end{cases}
\end{equation}
where we are using (\ref{astar}) in order to relate the on-shell Lagrangian to $a^*_d$.
Then, we observe that when $d$ is odd the $1/\delta'$ expansion contains the term
\begin{equation}\label{logbulk}
\int_{\delta'}^{\pi/2}d\theta'\theta'\frac{\cos^{d-1}\theta'}{\sin^{d+1}\theta'}=\frac{1}{(d-1)\delta'^{d-1}}+\ldots+(-1)^{\frac{d-1}{2}}\frac{1}{d}\log\left(\csc\delta'\right)+\ldots\, ,
\end{equation}
and therefore we should also include this logarithm in (\ref{CAuniv1}). Note that this term appears equally in both regularizations. On the other hand, when $d$ is even the bulk action contains a constant term:
\begin{equation}
\int_{\delta'}^{\pi/2}d\theta'\theta'\frac{\cos^{d-1}\theta'}{\sin^{d+1}\theta'}=\frac{1}{(d-1)\delta'^{d-1}}+\ldots+(-1)^{d/2}\frac{\pi}{2d}+\ldots\, .
\end{equation}
Note that on general grounds when $d$ is even the $1/\delta$ expansion of the complexity contains odd powers of the regulator like $1/\delta^{d-1-2k}$ and $\log(\delta)/\delta^{d-1-2k}$, so in principle no constants terms are expected to appear. The one in the bulk action is probably the only exception to this and therefore it could be universal. Putting it all together, we have found the following possible universal contributions to the complexity
\begin{equation}\label{CAuniv2}
\mathcal{C}^A_{\rm univ}=
\begin{cases}
(-1)^{\frac{d-2}{2}}2 a_d^*\quad &\text{if $d$ is even}\, ,\\
(-1)^{\frac{d-3}{2}}\frac{4a_d^*}{\pi}\log\left(\frac{\tilde L}{\delta}\right)+(-1)^{\frac{d-1}{2}}b_d \log\left(\frac{\tilde L}{\sqrt{\alpha_1\alpha_2}\delta}\right)\quad &\text{if $d$ is odd}\, .
\end{cases}
\end{equation}
In the case of odd boundary dimensions we are keeping the two logarithmic terms separated because they actually have different properties. This becomes evident if we allow the boundary of AdS to contain a $(d-1)$-sphere of arbitrary radius $R$. In the metric (\ref{FG}) this radius is fixed to be equal to the AdS scale $\tilde L$, but we may perform the change of variables $z=\hat z \tilde L/R$, $t=\hat t \tilde L/ R$, so that asymptotically the metric takes the form
\begin{equation}\label{GAdS2}
ds^2=\frac{\tilde L^2}{\hat z^2}\left(d\hat z^2-d\hat t^2+R^2d\Omega_{(d-1)}^2+\mathcal{O}(\hat z^2)\right)\, .
\end{equation}
The short-distance cutoff in this metric, $\hat\delta$, is related to the cutoff of the metric (\ref{FG}) according to $\delta=\hat\delta \tilde L/R$. Then, the result for the complexity in this metric could be obtained by replacing $\delta\rightarrow \hat \delta \tilde L/R$ everywhere in the previous expressions. However, there is a difference, because the natural choice of the null generators in this coordinate system now is $k_1=\alpha_1(d\hat z+d\hat t)$, $k_2=\alpha_2 (d\hat z-d\hat t)$. The product of these vectors evaluated at the joints yields $k_1\cdot k_2/2=\alpha_1\alpha_2 \hat\delta^2/\tilde L^2$, which is the argument of the logarithm appearing in the joint contribution. Hence, this logarithm is not changed at all and the universal contribution to the complexity reads
\begin{equation}\label{CAuniv3}
\mathcal{C}^A_{\rm univ}=
(-1)^{\frac{d-3}{2}}\frac{4a_d^*}{\pi}\log\left(\frac{R}{\hat\delta}\right)+(-1)^{\frac{d-1}{2}}b_d \log\left(\frac{\tilde L}{\sqrt{\alpha_1\alpha_2}\hat\delta}\right)\, .
\end{equation}
This shows that the first logarithm actually depends on the boundary geometry, while the scale in the second one really corresponds to the AdS curvature scale. The first one is controlled by the charge $a^*_d$ that appears in the holographic EE, while the second one depends on a different, probably previously unknown constant $b_d$. It has been argued that the combination $\tilde L/\sqrt{\alpha_1\alpha_2}$ would correspond to a new scale that appears in the microscopic rules defining the complexity in the CFT \cite{Jefferson:2017sdb,Carmi:2016wjl}.  This could give us a hint about the role of the constant $b_d$ in the CFT.

We have tried to argue that the boundary contributions will not affect this result, but of course a more rigorous computation taking into account the null boundaries is necessary. In addition, it is not clear what relevance one can asses to the universal terms that we have obtained due to the ambiguities present in the computation of the complexity. We have shown that these terms are at least independent of the regularization scheme,  but we still have the ambiguity in the parameters $\alpha_{1,2}$ which fix the normalization of the null generators.

\section{Discussion}\label{4}
In this work we have obtained the general form of the Lovelock action when the spacetime domain contains nonsmooth joints between spacelike or timelike segments --- the details are explained in section \ref{CLove}. The results here were obtained by making use of the smoothing method for type I joints and afterwards we showed that they can be extended to type II joints as well.  It would be interesting to try a direct proof by using the variational method. This is probably more challenging, but now that we know the general structure of these terms we have a very useful hint on how to tackle this computation. With these boundary and joint terms the action is additive, in the same sense as in Einstein gravity \cite{PhysRevD.50.4914,Lehner:2016vdi}. 

Let us also point out that we have only considered intersections of segments producing codimension 2 joints, but in general the boundary could contain higher-order joints --- vertices ---  and in general they will also yield a contribution. If we denote by $k$-vertex a codimension $k$ defect in the boundary --- the usual joints would be 2-vertices --- then we expect that they will contribute to the $n$-th Lovelock action if $k\le n$. For example, a 3-vertex does not have any effect in the Einstein-Hilbert action, but it does contribute to the Gauss-Bonnet one. These additional vertex terms could be computed by using similar techniques to those presented here.

As we have seen, there are two kind of contributions that appear when the boundary contains nonsmooth joints. The first one --- the term $\mathcal{F}_n$ in (\ref{LoveFull}) --- is related to the presence of a boundary in each segment of the boundary. This contribution is independent of the joint, and we can actually think of it as adding a certain total derivative to the usual boundary term (\ref{bdry}).  The form of this total derivative seems to depend on the decomposition used to describe the boundary of every segment though. The second piece is the actual corner contribution, since it depends on the angle $\Theta$ of the corner in the Euclidean case, or in the local Lorentz boost parameter $\eta$ in the Lorentzian one.  This term has a very appealing form, since it involves the integral along the joint of the Jacobson-Myers entropy density \cite{Jacobson:1993xs} weighted by the aforementioned parameter $\psi=\Theta, \eta$. If we write the Lagrangian density as $\mathcal{L}=\frac{1}{16\pi G}\left(R+\sum_n\lambda_n \mathcal{X}_{2n}\right)$ then the contribution of the joints reads
\begin{equation}
I_{\rm joint}=\frac{1}{2\pi}\int_{\mathcal{C}}d\sigma \psi \rho_{\rm JM}\, ,\quad \text{where}\quad \rho_{\rm JM}=\frac{1}{4 G}\left(1+\sum_n n\lambda_n \hat{\mathcal{X}}_{2(n-1)}\right)
\end{equation} 
This remarkable result could have relevant consequences. For example, in situations where $\psi$ is constant, this term is actually picking an entropy in the action. In particular, if $\mathcal{C}$ is the bifurcation surface of a black hole horizon, then the Jacobson-Myers entropy equals the Wald entropy \cite{Wald:1993nt}, and the joint contributes with a quantity proportional to the black hole entropy. A similar result ---though with an slightly different point of view--- was obtained in \cite{Neiman:2013ap}, where it was observed that the Jacobson-Myers entropy appears in the ``imaginary part'' of the action. In that context, the imaginary part of the action is precisely interpreted as the black hole entropy, but in our case this term contributes to the variational problem. Also, the Jacobson-Myers entropy coincides with the  functional for the holographic entanglement entropy in higher-derivative gravity \cite{Camps:2013zua,Dong:2013qoa}, so again in certain situations this term may yield something related to an entanglement entropy. 

In addition, we have shown that $I_{\rm joint}$ can be straightforwardly extended to the case of intersections of null boundaries just by using the same value for $\psi$ as in Einstein gravity. Although the appropriate surface terms for null boundaries are still unknown for Lovelock gravity (including the contributions $\mathcal{F}_n$), we have used this result in order to compute the ``logarithmic part'' of the holographic complexity of AdS, since the logarithmic contribution was argued to come only from the joints. We found that the higher-curvature corrections do not modify the leading divergence in this series, but they do contribute to the subleading terms. In particular, when $d$ is odd the series contains a ``universal term'' of the form $\mathcal{C}_A^{\rm univ}=(-1)^{\frac{d-1}{2}}b_d \log\left(\frac{\tilde L}{\sqrt{\alpha_1\alpha_2}\delta}\right)$. All of the Lovelock densities contribute to the constant $b_d$ --- see (\ref{Cd}) --- which could be a new universal quantity characterizing the dual CFT, analogous to $a^*_d$. If that were the case, the dependence of this ``charge'' on the Lovelock couplings could be relevant in order to identify a possible complexity model producing this constant from the CFT side. By computing the bulk action, we have seen that $a^*_d$ also plays a role in the complexity (\ref{CAuniv2}) --- a possible appearance of $a_d^*$ in the complexity in general theories was also anticipated in \cite{Bueno:2018xqc}. However, a careful inspection of the null boundary terms is still necessary in order to assess the validity of this result. In addition, we should be careful regarding the interpretation of possible universal terms, since there are some ambiguities in the complexity. In particular, it depends on the free parameters $\alpha_1$, $\alpha_2$ of the null generators. It has been argued that these parameters may be chosen in a way such that $\tilde L/\sqrt{\alpha_1\alpha_2}\equiv\ell$ is a new length scale which would appear in the UV physics of the CFT \cite{Carmi:2016wjl}. It could happen that this new scale is actually cutoff dependent, $\ell=\delta e^{\sigma}$, and such property would actually spoil the universal character of the logarithmic term proportional to $b_d$.  In any event, it seems clear that exploring the dependence of the complexity on the higher-curvature couplings could give us relevant information about this quantity. 

Although we have focused on computing the complexity of global AdS, it would also be interesting to estimate the subregion complexity \cite{Carmi:2016wjl} in Lovelock gravity. 
With the results at hands one could also try to re-compute the action growth for Lovelock black holes \cite{Cai:2016xho} by using the methods of \cite{Lehner:2016vdi, Carmi:2017jqz}, or the complexity of formation \cite{Chapman:2016hwi}, which has the advantage of being finite and independent of the normalization of the null generators. In particular, it would be interesting to explore if the full time dependence of the complexity in Lovelock black holes shares the same properties found in Einstein gravity \cite{Brown:2015lvg,Carmi:2017jqz}, where the asymptotic behavior of the complexity at late times is $\mathcal{C}_A\propto M t$. In the case of Einstein gravity the null surface terms vanish when we choose the null generators to be affinely parametrized \cite{Lehner:2016vdi}, so that all the contributions to the complexity come from the bulk action and the joints, which we also know for Lovelock gravity. However, it is unclear that the null surface terms will also vanish in this case. A more rigorous computation would require the analysis of the null boundary terms for Lovelock gravity, which are still unknown.

\acknowledgments
I would like to thank Rob Myers for guidance and advice during this project and Pablo Bueno, Dean Carmi, Robie Hennigar, Hugo Marrochio, Tom\'as Ort\'in and Shan-Ming Ruan for useful discussions and comments. The work of the author is funded by Fundaci\'on la Caixa through a ``la Caixa - Severo Ochoa" international pre-doctoral grant. This work was also supported by Perimeter Institute through the ``Visiting Graduate Fellows" program, by the MINECO/FEDER, UE grant FPA2015-66793-P and by the Spanish Research Agency (Agencia Estatal de Investigaci\'on) through the grant IFT Centro de Excelencia Severo Ochoa SEV-2016- 0597. Research at Perimeter Institute is supported by the Government of Canada through the Department of Innovation, Science and Economic Development and by the Province of Ontario through the Ministry of Research, Innovation and Science.

\appendix

\section{Variation of the Lovelock action}
First, after some algebra in (\ref{varf}), we are able to express the boundary term as
\begin{equation}
\begin{aligned}
n_{\mu}\dslash  v_n^{\mu}=&\frac{n}{2^{n-1}}\left(n_{\alpha}h^{\sigma\beta}\delta \Gamma_{\nu\sigma}^{\ \ \ \alpha}-n^{\sigma}\delta \Gamma_{\nu\sigma}^{\ \ \ \beta}\right)h^{\nu \sigma_1\dots\sigma_{2n-2}}_{\beta \lambda_1\dots \lambda_{2n-2}}R_{\sigma_1\sigma_2}^{\lambda_1\lambda_2}\cdots R_{\sigma_{2n-3}\sigma_{2n-2}}^{\lambda_{2n-3}\lambda_{2n-2}}\\
&-\frac{n(n-1)}{2^{n-2}}\delta \Gamma_{\nu\sigma}^{\ \ \ \alpha}h^{\sigma\beta}h^{\nu \sigma_1\sigma_2\dots\sigma_{2n-2}}_{\beta \alpha \lambda_2\dots \lambda_{2n-2}}n_{\lambda_1}R_{\sigma_1\sigma_2}^{\lambda_1\lambda_2}\cdots R_{\sigma_{2n-3}\sigma_{2n-2}}^{\lambda_{2n-3}\lambda_{2n-2}}\, .
\end{aligned}
\end{equation}
Now, we can write these expressions in terms of intrinsic quantities by using the Gauss-Codazzi equations:
\begin{eqnarray}
e^{i}_{\alpha}e^{j}_{\beta}e^{\mu}_{k}e^{\nu}_{l}R^{\alpha\beta}_{\mu\nu}&=&\mathcal{R}^{ij}_{kl}-2\epsilon K^{i}_{[k}K^{j}_{l]}\, ,\\
n_{\lambda}e_{i}^{\sigma}e_{j}^{\alpha}e_{k}^{\beta}R^{\lambda}_{\ \ \sigma\alpha\beta}&=&-2D_{[i}K_{j]k}\, ,
\end{eqnarray}
where $D_i$ is the covariant derivative on the boundary. Then, we can write

\begin{equation}
\begin{aligned}
n_{\mu}\dslash v_n^{\mu}=&\frac{n}{2^{n-1}}\left(n_{\alpha}h^{\sigma\beta}\delta \Gamma_{\nu\sigma}^{\ \ \ \alpha}-n^{\sigma}\delta \Gamma_{\nu\sigma}^{\ \ \ \beta}\right)e_{\beta}^{j}e^{\nu}_{i}\delta^{i i_1\dots i_{2n-2}}_{j j_1\dots j_{2n-2}}\left(\mathcal{R}^{j_1 j_2}_{i_1i_2}-2\epsilon K^{j_1}_{i_1}K^{j_2}_{i_2}\right)\cdots \\ &\left(\mathcal{R}^{j_{2n-3}j_{2n-2}}_{i_{2n-3}i_{2n-2}}-2\epsilon K^{j_{2n-3}}_{i_{2n-3}}K^{j_{2n-2}}_{i_{2n-2}}\right)\\
&+\frac{n(n-1)}{2^{n-3}}\delta \Gamma_{\nu\sigma}^{\ \ \ \alpha}e^{\sigma k} e^{\nu}_{i}e_{\alpha}^{j_1}\delta^{i i_1 i_2\dots i_{2n-2}}_{k j_1 j_2\dots j_{2n-2}}D_{i_1}K_{i_2}^{j_2}\left(\mathcal{R}^{j_3 j_4}_{i_3i_4}-2\epsilon K^{j_3}_{i_3}K^{j_4}_{i_4}\right)\cdots \\
&\left(\mathcal{R}^{j_{2n-3}j_{2n-2}}_{i_{2n-3}i_{2n-2}}-2\epsilon K^{j_{2n-3}}_{i_{2n-3}}K^{j_{2n-2}}_{i_{2n-2}}\right)\, .
\end{aligned}
\end{equation}
Now let us compute $\delta I_{\rm bdry}^{(n)}$. For simplicity, we will not keep trace of terms proportional to $\delta h_{ij}$, since eventually we will set $\delta h_{ij}=0$.  Hence:
\begin{equation}
 \delta I_{\rm bdry}^{(n)}= \int_{\partial \mathcal{M}}d^{d}x\sqrt{|h|}\left(\delta \mathcal{Q}_{n}+\mathcal{O}(\delta h)\right)\, .
\end{equation}
The variation of $\mathcal{Q}_n$ can be separated in two components: variations through the extrinsic curvature and variations through the intrinsic one. The former reads:
\begin{equation}
\delta_{K}\mathcal{Q}_n=\frac{n}{2^{n-2}}\delta^{i_1\dots i_{2n-1}}_{j_1\dots j_{2n-1}}\delta K^{j_1}_{i_1}\left(\mathcal{R}^{j_2 j_3}_{i_2i_3}-2\epsilon K^{j_2}_{i_2}K^{j_3}_{i_3}\right)\cdots \left(\mathcal{R}^{j_{2n-2}j_{2n-1}}_{i_{2n-2}i_{2n-1}}-2\epsilon K^{j_{2n-2}}_{i_{2n-2}}K^{j_{2n-1}}_{i_{2n-1}}\right)\, .
\end{equation}
Note the funny effect of ``canceling'' the integration. On the other hand we have to consider variations with respect to the intrinsic curvature. In this case, we know that
\begin{equation}
\delta \mathcal{R}^{i}_{\  jkl}=2 D_{[k}\delta \hat\Gamma_{l]j}^{\ \ i}\, ,
\end{equation}
where $ \hat\Gamma_{lj}^{\ \ i}$ is the Levi-Civita connection of the induced metric $h_{ij}$. Hence, whenever $\delta \mathcal{R}^{i}_{\  jkl}$ appear we can integrate by parts, and by using the Bianchi identity $D_{[i}\mathcal{R}^{jk}_{lm]}=0$ we obtain the following result:
\begin{equation}
\begin{aligned}
\delta_{R}\mathcal{Q}_n=&D_ i F^i-\frac{n(n-1)}{2^{n-3}}\delta \hat\Gamma_{i l}^{\ \ \ j_1}h^{l k} \delta^{i i_1 i_2\dots i_{2n-2}}_{k j_1 j_2\dots j_{2n-2}}D_{i_1}K_{i_2}^{j_2}\left(\mathcal{R}^{j_3 j_4}_{i_3i_4}-2\epsilon K^{j_3}_{i_3}K^{j_4}_{i_4}\right)\cdots \\
&\left(\mathcal{R}^{j_{2n-3}j_{2n-2}}_{i_{2n-3}i_{2n-2}}-2\epsilon K^{j_{2n-3}}_{i_{2n-3}}K^{j_{2n-2}}_{i_{2n-2}}\right)+\mathcal{O}(\delta h)\,,
\end{aligned}
\end{equation}
where

\begin{equation}
\begin{aligned}
F^i=&-\frac{n(n-1)}{2^{n-3}}\int_{0}^{1}dt \delta^{i i_1 i_2\dots i_{2n-2}}_{j j_1 j_2\dots j_{2n-2}} K^{j}_{i1} h^{j_2 l}\delta \hat\Gamma_{i_2 l}^{\ \ \ j_1}\left(\mathcal{R}^{j_3 j_4}_{i_3i_4}-2\epsilon t^2 K^{j_3}_{i_3}K^{j_4}_{i_4}\right)\cdots \\
&\left(\mathcal{R}^{j_{2n-3}j_{2n-2}}_{i_{2n-3}i_{2n-2}}-2\epsilon t^2 K^{j_{2n-3}}_{i_{2n-3}}K^{j_{2n-2}}_{i_{2n-2}}\right)\,.
\end{aligned}
\end{equation}
The total variation of $\mathcal{Q}_n$ is then $\delta \mathcal{Q}_n=\delta_{K} \mathcal{Q}_n+\delta_{R} \mathcal{Q}_n$. It remains to compute $\delta K^{j}_{i}$. We can use the results in \cite{Lehner:2016vdi}:
\begin{equation}
\delta K^{j}_{i}=\frac{1}{2}\delta h^{jk} K_{ki}-\frac{1}{2}\epsilon D_i \delta A^j+\frac{1}{2}e_{i}^{\nu}e^{j}_{\beta}n^{\sigma}\delta \Gamma_{\nu\sigma}^{\ \ \ \beta}-\frac{1}{2}e_{i}^{\nu}e^{j}_{\beta}n_{\alpha}h^{\sigma\beta}\delta \Gamma_{\nu\sigma}^{\ \ \ \alpha}\, ,
\end{equation}
where $\dslash  A^j=-\epsilon e^{j}_{\alpha}n_{\beta}\delta g^{\alpha\beta}$. Then, if we add up all the results we obtain:
\begin{equation}
\begin{aligned}
&n_{\mu}\dslash  v^{\mu}+\delta \mathcal{Q}_n=-\epsilon\frac{n}{2^{n-1}}\delta^{i_1\dots i_{2n-1}}_{j_1\dots j_{2n-1}}D_{i_1} \dslash  A^{j_1}\left(\mathcal{R}^{j_2 j_3}_{i_2i_3}-2\epsilon K^{j_2}_{i_2}K^{j_3}_{i_3}\right)\cdots \left(\mathcal{R}^{j_{2n-2}j_{2n-1}}_{i_{2n-2}i_{2n-1}}-2\epsilon K^{j_{2n-2}}_{i_{2n-2}}K^{j_{2n-1}}_{i_{2n-1}}\right)\\
&+\frac{n(n-1)}{2^{n-3}}\left(\delta \Gamma_{\nu\sigma}^{\ \ \alpha}e^{\sigma}_{l} e^{\nu}_{i}e_{\alpha}^{j_1}-\delta \hat\Gamma_{i l}^{\ \ j_1}\right)h^{lk}\delta^{i i_1 i_2\dots i_{2n-2}}_{k j_1 j_2\dots j_{2n-2}}D_{i_1}K_{i_2}^{j_2}\left(\mathcal{R}^{j_3 j_4}_{i_3i_4}-2\epsilon K^{j_3}_{i_3}K^{j_4}_{i_4}\right)\cdots \\
&\left(\mathcal{R}^{j_{2n-3}j_{2n-2}}_{i_{2n-3}i_{2n-2}}-2\epsilon K^{j_{2n-3}}_{i_{2n-3}}K^{j_{2n-2}}_{i_{2n-2}}\right)+D_i F^i+\mathcal{O}(\delta h)\, .
\end{aligned}
\end{equation}
Now, in the first term we integrate by parts, and after using the Bianchi identity we get
\begin{equation}
\begin{aligned}
&n_{\mu}\dslash  v^{\mu}+\delta \mathcal{Q}_n=D_i (F^i+G^i)+\mathcal{O}(\delta h)\\
&+\frac{n(n-1)}{2^{n-3}}\left(\delta \Gamma_{\nu\sigma}^{\ \ \alpha}e^{\sigma}_{l} e^{\nu}_{i}e_{\alpha}^{j_1}-\delta \hat\Gamma_{i l}^{\ \ j_1}-K_{i l}\dslash  A^{j_1}\right)h^{lk}\delta^{i i_1 i_2\dots i_{2n-2}}_{k j_1 j_2\dots j_{2n-2}}D_{i_1}K_{i_2}^{j_2}\left(\mathcal{R}^{j_3 j_4}_{i_3i_4}-2\epsilon K^{j_3}_{i_3}K^{j_4}_{i_4}\right)\cdots \\
&\left(\mathcal{R}^{j_{2n-3}j_{2n-2}}_{i_{2n-3}i_{2n-2}}-2\epsilon K^{j_{2n-3}}_{i_{2n-3}}K^{j_{2n-2}}_{i_{2n-2}}\right)\, ,
\end{aligned}
\end{equation}
where
\begin{equation}
G^{i_1}=-\epsilon\frac{n}{2^{n-1}}\delta^{i_1 \dots i_{2n-1}}_{j_1\dots j_{2n-1}} \dslash  A^{j_1}\left(\mathcal{R}^{j_2 j_3}_{i_2i_3}-2\epsilon K^{j_2}_{i_2}K^{j_3}_{i_3}\right)\cdots \left(\mathcal{R}^{j_{2n-2}j_{2n-1}}_{i_{2n-2}i_{2n-1}}-2\epsilon K^{j_{2n-2}}_{i_{2n-2}}K^{j_{2n-1}}_{i_{2n-1}}\right)\, .
\end{equation}
Finally, using that $g_{\mu\nu}=e_{\mu}^{i}e_{\nu}^{j}h_{ij}+\epsilon n_{\mu}n_{\nu}$ it is possible to show that $\delta \Gamma_{\nu\sigma}^{\ \ \alpha}e^{\sigma}_{l} e^{\nu}_{i}e_{\alpha}^{j_1}-\delta \hat\Gamma_{i l}^{\ \ j_1}-K_{i l}\dslash  A^{j_1}=0$. 

Hence, we have the following result:

\begin{equation}
\int_{\partial \mathcal{M}} d\Sigma_{\mu}\dslash v^{\mu}_{n}+  \delta I_{\rm bdry}^{(n)}=\int_{\partial\mathcal{M}} d^{d}x\sqrt{|h|}\left( T^{ij}\delta h_{ij}+D_i \dslash H^i\right)\, ,
\end{equation}
for certain $T^{ij}$ that we do not worry about and where $\dslash  H^i=F^i+G^i$.

\section{Example: Gauss-Bonnet theorem in $D=4$}
According to our results, the Euler characteristic of a 4-dimensional manifold with (non-smooth) boundary must be given by
\begin{equation}\label{GBTH}
\begin{aligned}
\mathcal{X}(\mathcal{M})=&\frac{1}{32\pi^2}\int_{\mathcal{M}}d^{4}x\sqrt{g}\mathcal{X}_4+ \frac{1}{16\pi^2}\int_{\partial \mathcal{M}}d^{3}x\sqrt{h}\delta^{i_1i_2i_3}_{j_1j_2j_3}K^{j_1}_{i_1}\left(\mathcal{R}^{j_2 j_3}_{i_2i_3}-\frac{2}{3}K^{j_2}_{i_2}K^{j_3}_{i_3}\right)\\
&+\frac{1}{8\pi^2}\sum_l\int_{\mathcal{C}_l}d\sigma \left[\Theta\hat{\mathcal{R}}+2L^{[A}_{1\, A}Q^{B]}_{1\, B}+2L^{[A}_{2\, A}Q^{B]}_{2\, B}\right]\, .
\end{aligned}
\end{equation}
Let us check this result by evaluating it in the following deformed 4-dimensional cylinder embedded on flat space
\begin{equation}
\mathcal{M}=\big\{(x,y,z,u)\in \mathbb{R}^4/ x^2+y^2+z^2\le1\, , 0\le u\le f(z)\big\}\, ,
\end{equation}
where $f$ is an arbitrary positive, differentiable function. Since this manifold is topologically $B_4$, the Euler characteristic is $1$, and (\ref{GBTH}) should give us this result for any function $f$. The bulk piece vanishes because we are working in flat space. In the boundary term it is useful to apply the Gauss-Codazzi equations, so that we get
\begin{equation}
\mathcal{R}^{ij}_{kl}=R^{ij}_{kl}+2K^{i}_{[k}K^{j}_{l]}=2K^{i}_{[k}K^{j}_{l]}\, ,
\end{equation}
where we again use that the curvature $R^{ij}_{kl}$ vanishes. Then, the boundary contribution only contains the combination $K^{[i_1}_{i_1}K^{i_2}_{i_2}K^{i_3]}_{i_3}\propto \det(K)$. Therefore, the extrinsic curvature only contributes when it has rank 3. To determine $K_{ij}$ for the different pieces of the boundary, let us write the Euclidean space in cylindrical  coordinates
\begin{equation}
ds^2=dr^2+r^2(d\psi^2+\sin^2\psi d\phi^2)+du^2\, .
\end{equation}
Then there are three pieces: the top $(\mathcal{T})$ $u=f(z)$, the bottom $(\mathcal{B})$ $u=0$ and the side $(\mathcal{S})$ of the cylinder $r=1$. The normal to $\mathcal{B}$ is $n=-du$ and the extrinsic curvature is vanishing. In $\mathcal{S}$ we have $n=dr$ and $K=(d\psi^2+\sin^2\psi d\phi^2)$ has rank 2, so the contribution is zero. Finally, in the top $\mathcal{T}$ the normal is $n=\frac{du-f'(z) dz}{\sqrt{1+f'(z)^2}}$ and its extrinsic curvature has rank 1, so it does not contribute either. Therefore, all the contribution comes from the joints. Let us start with the joint $\mathcal{C}_1=\mathcal{S}\cap \mathcal{B}$. This joint is defined by the intersection of the surfaces $r=1$ and $u=0$ and therefore it is a 2-sphere, which implies that the induced curvature takes the value $\hat{\mathcal{R}}=2$. We have the following system of adapted normals in each piece of boundary: the normals adapted to $\mathcal{B}$ are $(n_1, s_1)=(-du,dr)$ while those coming from $\mathcal{S}$ are $(n_2, s_2)=(dr, -du)$. Obviously, the change in the normal is $\Theta=\arccos(n_1\cdot n_2)=\pi/2$. On the other hand, the extrinsic curvatures are $L_1=Q_2=0$, $Q_1=L_2=(d\psi^2+\sin^2\psi d\phi^2)$. Therefore, the contribution from this joint is 
\begin{equation}
\frac{1}{8\pi^2}\int_{\mathcal{C}_1}d\sigma \left[\Theta\hat{\mathcal{R}}+2L^{[A}_{1\, A}Q^{B]}_{1\, B}+2L^{[A}_{2\, A}Q^{B]}_{2\, B}\right] =\frac{1}{8\pi^2}\int_{\mathcal{C}_1}d\sigma \pi=\frac{1}{2}\, .
\end{equation}
Then, let us finally consider the joint $\mathcal{C}_2=\mathcal{S}\cap\mathcal{T}$ between the surfaces $r=1$ and $u=f(z)=f(r \cos\psi)$. We have the following normals in each boundary
\begin{equation}
n_1=dr\, ,\quad n_2=\frac{du-f' dr \cos\psi+f' r \sin \psi d\psi }{\sqrt{1+f'^2}}\, ,
\end{equation}
where we are writing in short $f'\equiv f'(r\cos\psi)$. From these normals to the boundaries we construct $s_1$ and $s_2$ such that they unitary and tangent to their respective boundary and pointing outwards. This is, $n_i\cdot s_i=0$, $s_i^2=1$, $i=1,2$. We find
\begin{eqnarray}
s_1&=&\frac{du+f' d\psi r \sin\psi }{\sqrt{1+\sin^2\psi f'^2}}\, ,\\
s_2&=&\frac{ f' \cos\psi du+(1+\sin^2\psi f'^2)dr+r\cos\psi\sin\psi f'^2 d\psi}{\sqrt{(1+f'^2)(1+\sin^2\psi f'^2)}}\, .
\end{eqnarray}
These expressions have to be evaluated at $r=1$ but it is useful to keep trace of all the coordinates in order to compute the extrinsic curvatures $L_i=Dn_i$, $Q_i=Ds_i$. They read
\begin{eqnarray}
L_1&=&d\psi^2+\sin^2\psi d\phi^2\, ,\\
Q_1&=&\frac{-\sin^2\psi f'' d\psi^2}{\sqrt{1+f'^2}}\, ,\\
L_2&=&\frac{\left(\cos\psi f'-\sin^2\psi f''\right)d\psi^2+\cos\psi\sin^2\psi f'd\phi^2}{\sqrt{1+\sin^2\psi f'^2}}\, ,\\
Q_2&=&\frac{\left(1+f'^2-\cos\psi\sin^2\psi f'f''\right)d\psi^2-\sin^2\psi\left(1+f'^2\right)d\phi^2}{\sqrt{(1+f'^2)(1+\sin^2\psi f'^2)}}
\end{eqnarray}
We also find the angle $\Theta$ in which the normal changes:
\begin{equation}
\Theta=\arccos(n_1\cdot n_2)=\arccos\left[\frac{-\cos\psi f'(\cos\psi)}{\sqrt{1+f'(\cos\psi)^2}}\right]\, .
\end{equation}
On the other hand, the induced metric reads
\begin{equation}
^{(2)}ds^2=\left(1+\sin^2\psi f'(\cos\psi)^2\right)d\psi^2+\sin^2\psi d\phi^2\, .
\end{equation}
Then we get the Ricci scalar and the combinations of extrinsic curvatures
\begin{eqnarray}
\mathcal{R}&=&\frac{2\left(1+f'^2-\cos\psi\sin^2\psi f'f''\right)}{(1+\sin^2\psi f'^2)^2}\, ,\\
2L^{[A}_{1\, A}Q^{B]}_{1\, B}&=&\frac{2\cos\psi f'-\sin^2\psi f''}{(1+\sin^2\psi f'^2)^{3/2}}\, ,\\
2L^{[A}_{2\, A}Q^{B]}_{2\, B}&=&\frac{-\sin^2\psi f''}{(1+\sin^2\psi f'^2)^{3/2}}\, .
\end{eqnarray}
Putting it all together, the Euler characteristic reads
\begin{equation}
\begin{aligned}
\mathcal{X}(\mathcal{M})&=\frac{1}{2}+\frac{1}{8\pi^2}\int_{\mathcal{C}_2}d\sigma \left[\Theta\hat{\mathcal{R}}+2L^{[A}_{1\, A}Q^{B]}_{1\, B}+2L^{[A}_{2\, A}Q^{B]}_{2\, B}\right]\\
&=\frac{1}{2}+\frac{1}{4\pi}\int_{-1}^{1}dx\left[\frac{2 \left(1+f'^2+x \left(x^2-1\right) f' f''\right) \arccos \left(\frac{-x f'}{\sqrt{f'^2+1}}\right)}{\left(1+\left(1-x^2\right) f'^2\right)^{3/2}}+\frac{2 \left(\left(x^2-1\right) f''+x f'\right)}{1+\left(1-x^2\right) f'^2}\right]\, ,
\end{aligned}
\end{equation}
where we introduced $x=\cos\psi$, and $f'\equiv f'(x)$, $f''\equiv f''(x)$. The integrand is a total derivative, and thus can be integrated and we obtain
\begin{equation}
\begin{aligned}
\mathcal{X}(\mathcal{M})&=\frac{1}{2}+\frac{1}{4\pi}\left[\frac{2 x \arccos \left(\frac{-x f'(x)}{\sqrt{f'(x)^2+1}}\right)}{\sqrt{1+\left(1-x^2\right) f'(x)^2}}-2 \arctan\left(f'(x)\right)\right]_{x=-1}^{x=1}\\
&=\frac{1}{2}+\frac{1}{2\pi}\left[\arccos\left(\frac{a}{\sqrt{a^2+1}}\right)+\arctan(a)+\arccos\left(\frac{-b}{\sqrt{b^2+1}}\right)+\arctan(-b)\right]\, ,
\end{aligned}
\end{equation}
where $a=f'(-1)$ and $b=f'(1)$. But now, $\arccos\left(\frac{a}{\sqrt{a^2+1}}\right)+\arctan(a)=\pi/2$  $\forall a\in\mathbb{R}$. Therefore, we have shown that $\mathcal{X}(\mathcal{M})=1$ and the Gauss-Bonnet theorem works when we include the joint terms. In particular note that the terms involving the extrinsic curvatures are necessary in order to obtain a topological result.


\renewcommand{\leftmark}{\MakeUppercase{Bibliography}}
\phantomsection
\bibliographystyle{JHEP}
\bibliography{Gravities}
\label{biblio}

\end{document}